\documentclass[nofootinbib,aip,amsmath,amssymb,reprint]{revtex4-1}

\usepackage{graphicx}
\usepackage{hyperref}
\graphicspath{{./figures/}}
\usepackage{dcolumn}
\usepackage{bm}
\usepackage[utf8]{inputenc}
\usepackage[T1]{fontenc}
\usepackage{mathptmx}

\usepackage{xcolor}

\begin{document}

\title{Spatio-temporal correlations in 3D homogeneous isotropic turbulence} 

\author{A. Gorbunova}
\affiliation{Université Grenoble Alpes, Centre National de la Recherche Scientifique, Laboratoire de Physique et Mod\'elisation des Milieux Condens\'es, 38000 Grenoble, France}
\affiliation{Université Grenoble Alpes, Centre National de la Recherche Scientifique, Laboratoire des Ecoulements G\'eophysiques et Industriels, 38000 Grenoble, France}

\author{G. Balarac}
\affiliation{Université Grenoble Alpes, Centre National de la Recherche Scientifique, Laboratoire des Ecoulements G\'eophysiques et Industriels, 38000 Grenoble, France}
\affiliation{Institut Universitaire de France, 1 rue Descartes, 75000 Paris, France}

\author{L. Canet}
\affiliation{Université Grenoble Alpes, Centre National de la Recherche Scientifique, Laboratoire de Physique et Mod\'elisation des Milieux Condens\'es, 38000 Grenoble, France}
\affiliation{Institut Universitaire de France, 1 rue Descartes, 75000 Paris, France}

\author{G. Eyink}
\affiliation{Department of Applied Mathematics \& Statistics, The Johns Hopkins University, Baltimore, MD, USA, 21218}
\affiliation{Department of Physics \& Astronomy, The Johns Hopkins University, Baltimore, MD, USA, 21218}

\author{V. Rossetto}
\affiliation{Université Grenoble Alpes, Centre National de la Recherche Scientifique, Laboratoire de Physique et Mod\'elisation des Milieux Condens\'es, 38000 Grenoble, France}

\date{\today}

\begin{abstract}
We use Direct Numerical Simulations (DNS) of the forced
Navier-Stokes equation for a 3-dimensional incompressible fluid 
in order to test recent theoretical predictions.
We study the two- and three-point spatio-temporal correlation functions of the velocity field
in stationary, isotropic and homogeneous turbulence.
We compare our numerical results to the predictions from the Functional 
Renormalization Group (FRG) which were obtained in the large wavenumber limit.
DNS are performed at various Reynolds numbers
and the correlations are
analyzed in different time regimes focusing on the large wavenumbers.
At small time delays, we find that the two-point correlation function decays as
a Gaussian in the variable $kt$ where $k$ is the wavenumber and~$t$ 
the time delay.
The three-point correlation function, determined from the time-dependent
advection-velocity correlations,  also  follows a Gaussian decay at small $t$
with the same prefactor as the one of the two-point function. These  behaviors
are in precise agreement with the FRG results, and can be simply understood as
a consequence of  sweeping.  At large time delays, the FRG predicts a  crossover to an exponential in $k^2 t$, which we were not able to resolve in our
simulations.  However, we analyze the two-point spatio-temporal correlations of
the modulus of the velocity, and show that they exhibit this crossover from a
Gaussian to an exponential decay, although we lack of a theoretical
understanding in this case. This intriguing phenomenon calls for further
theoretical investigation.
\end{abstract}

\maketitle

\section{Introduction}
\label{sec:intro}

\par
Characterizing the statistical properties of a turbulent flow is one of the main challenges to achieve a complete theoretical understanding of turbulence.
Space-time correlations are at the heart of statistical theories of turbulence,
and have been studied and modeled for many decades, both in the Eulerian and Lagrangian frameworks
\cite{wallaceSpacetimeCorrelationsTurbulent2014, heSpaceTimeCorrelationsDynamic2017}.
One of the earliest insights was provided by Taylor's celebrated analysis of single particle dispersion 
by an isotropic turbulent flow \cite{taylor1922diffusion}. 
The understanding of the behavior of turbulent fluctuations both in space and
time is  essential for many problems in fluid mechanics where the multiscale
temporal dynamics plays a key role, such as particle-laden turbulence,
propagation of waves in a turbulent medium
or turbulence-generated noise in compressible flows 
\cite{heSpaceTimeCorrelationsDynamic2017}.
Space-time correlations are also central for many closure schemes, such as the
direct-interaction Approximation (DIA) elaborated by Kraichnan
\cite{kraichnanKolmogorovHypothesesEulerian1964},  
or the eddy damped quasi-normal Markovian (EDQNM) approximation
\cite{lesieurAnalyticalTheoriesStochastic2008}. The accurate description of the spatio-temporal correlations is crucial for developing time-accurate
large-eddy simulation (LES) turbulence models, as well as for the analysis of
experimental data, for example, to assess the validity and corrections to the Taylor's frozen flow model used for time-to-space conversion of measurements.

\par A fundamental ingredient to understand the temporal behavior of turbulent flows in the Eulerian frame is the sweeping effect,  which was early 
identified in References \onlinecite{heisenberg1948,
kraichnanStructureIsotropicTurbulence1959,
kraichnanKolmogorovHypothesesEulerian1964,
tennekesEulerianLagrangianTime1975}. The random sweeping effect results from
the random advection of small-scale velocities by the large-scale
energy-containing eddies, 
even in the absence of mean flow.
This random sweeping was anticipated to 
induce a Gaussian decay in the variable $tk$, where $k$ is the wavenumber and
$t$ the time delay, of the two-point correlations of the Eulerian velocity
field, based on simplified models of advection
\cite{kraichnanKolmogorovHypothesesEulerian1964}.
However, at the theoretical level, the effect of sweeping also induces, 
in the original formulation of DIA,
a $k^{-3/2}$ decay of the energy spectrum in the
inertial range instead of the Kolmogorov $k^{-5/3}$ scaling. This led Kraichnan
to a complete reformulation of his theory 
using Lagrangian space-time correlations instead of Eulerian ones. 
The dependence of the two-point correlation function in 
$k^2 t^2$ predicted from sweeping has
been  observed and confirmed in numerous numerical simulations~\citep{orszag1972, sanadaRandomSweepingEffect1992, chenSweepingDecorrelationIsotropic1989, heComputationSpacetimeCorrelations2004, favierSpaceTimeCorrelations2010,canetSpatiotemporalVelocityvelocityCorrelation2017}
and also in experiments~\citep{poulainDynamicsSpatialFourier2006}. 
A notable consequence of this dependence in the product $kt$
is that the frequency energy spectrum of Eulerian velocities
exhibits a $\omega^{-5/3}$ decay, instead of the $\omega^{-2}$  expected from
K41 scaling \cite{chevillard2005}.  \par The random sweeping hypothesis is also
a part of the elliptic approximation that provides a model for spatio-temporal
correlation in turbulent shear flows \cite{He2006ellipticmodel} combining the
decorrelation effect of the sweeping by large scales and the convection by the
mean flow, and provides a correction to Taylor frozen-flow model. The elliptic
approximation model has been tested in numerical simulations and experimental
measurements in Rayleigh-Bénard convection flows
\cite{heSpaceTimeCorrelationsDynamic2017}. In Ref.
\onlinecite{wilczekWavenumberFrequencySpectrum2012} a model of spatio-temporal
spectrum of turbulence is proposed in the presence of a mean flow
\cite{kraichnanKolmogorovHypothesesEulerian1964} departing from the Kraichnan's advection problem, 
which is consistent with the elliptical model.  
Fewer studies address multi-point correlations, although they are used as
part of closure models \cite{moninStatisticalFluidMechanics2013}. An expression
for the three-point correlation function in a specific wave-vector and time
configuration was obtained within the DIA 
\cite{kraichnanStructureIsotropicTurbulence1959, kraichnanKolmogorovHypothesesEulerian1964}, 
and multi-point correlation functions were studied numerically in Ref.
\onlinecite{biferale2011}.  

\par Although the random sweeping effect is 
phenomenologically known
for a long time, and the models based on it provide satisfactory descriptions,
the theoretical justification of the hypothesis of random sweeping directly
from the Navier-Stokes equation has remained a challenging task. The
application of the renormalization group approach to  turbulence developed by
Yakhot et al. led to the conclusion that the sweeping effect on space-time
correlations must be small\cite{yakhotSpaceTimeCorrelations1989}, which is not
in agreement with the $\omega^{-5/3}$ Eulerian spectrum. This result and its
validity are discussed in the Ref.
\onlinecite{chenSweepingDecorrelationIsotropic1989}. In another work
\cite{drivasLargescaleSweepingSmallscale2017} the effect of the random sweeping
was estimated with the use of equations of band-passed velocity advected by a
large scale velocity. This work demonstrated that the random
sweeping plays a dominant role in the
Navier-Stokes dynamics at small scales.
\par
Recently, a theoretical progress has been achieved using Functional
Renormalization Group (FRG), which has
yielded the general form of any multi-point
correlation (and response) function in the limit of large
wave-numbers. These expressions are established in the Eulerian frame, in a rigorous and systematic way.
For the two-point space-time correlations, the Gaussian decay in $tk$ is recovered for small time delays $t$,
while a crossover to a slower exponential decay in $t$ is predicted at large time delays.
Similar results are obtained for any generic correlations involving 
an arbitrary number of space-time points.
While the Gaussian regime is known to originate from sweeping, the
exponential large-delay regime was not yet predicted. 
We show in this work that this behavior can also be derived from the
original Taylor and Kraichnan's arguments, which provide a clear physical
interpretation of this result.

\par
The aim of this work is to make
precision tests of the FRG results using Direct Numerical Simulations (DNS)
of the forced Navier-Stokes equation. We analyze the two-point and three-point
correlations and 
accurately confirm the FRG prediction in the small-time
regime. Even though the
long-time regime remains elusive in the simulations
data due to the weakness of the signal amplitude in this regime and the lack of
statistics, we unveil a very similar crossover from a
Gaussian to an exponential decay in the correlations
of the modulus of the velocity field. However, this
observation lacks a theoretical explanation so far.

\par The paper is organized as follows.
 In Sec.~\ref{sec:theory}, we briefly introduce the functional and
nonperturbative renormalization group (FRG) framework and
review the theoretical predictions stemming from it on the time dependence of
multi-point  correlation functions.  We also provide a heuristic argument
allowing one to grasp the physical content of these results.
We present in Sec.~\ref{sec:DNS}  the results of our DNS analysis.
We analyze the small delay regime of the two-point correlation
functions in the Sec.~\ref{sec:2point} and that of the three-point
correlation function in the Sec.~\ref{sec:3point}.
The temporal behavior of the two-point correlation of the modulus of the
velocity is discussed in the Sec.~\ref{sec:2point-mod}.

\section{Theoretical framework}
\label{sec:theory}

\subsection{Theoretical results from functional renormalization group}
\label{sec:FRG}

\par The FRG is
a versatile method well-developed since the early 
1990's and used in a wide
range of applications, both in high-energy physics (quantum gravity and QCD),
condensed matter, 
quantum many-particle systems and statistical mechanics,
including disordered and nonequilibrium problems
(see References
\onlinecite{berges2002,kopietz2010,delamotte2012,dupuis2020} for reviews). 
This method has been employed in particular to study
the incompressible 3D Navier-Stokes equation in several works
\cite{tomassini1997,mejiamonasterio2012,canet2016,canetSpatiotemporalVelocityvelocityCorrelation2017,tarpinBreakingScaleInvariance2018,tarpin2019}.
We here focus on a recent result concerning the spatio-temporal dependence of
multi-point correlation functions of the turbulent velocity field in
homogeneous, isotropic and stationary conditions. The detailed derivation of
the theoretical results can be found in Ref.
\onlinecite{tarpinBreakingScaleInvariance2018}; it
relies on an expansion at
large wavenumbers of the exact FRG flow equations. 
The field theory arising from the stochastically forced Navier-Stokes equation
 possesses extended symmetries (in particular the time-dependent Galilean
symmetry) which allow one to obtain the exact leading term of this expansion.
We give below the ensuing expressions,
before providing their intuitive physical interpretation in the 
Sec.~\ref{sec:heuristic}.
  
\par  We are first interested in the two-point correlation function of the
velocity expressed in
the time-delay$-$wavevector mixed
coordinates $(t,\vec k)$, defined as
\begin{eqnarray}
 C^{(2)} (t, \vec{k}) &\equiv \text{FT} \left[ \left<  u_i(t_0, \vec{r}_0) u_i(t_0 + t, \vec{r}_0 + \vec{r}) \right> \right]  \nonumber \\  
	&= \left< \hat{u}_i(t_0, \vec{k}) \hat{u}^*_i(t_0 + t, \vec{k}) \right> 
	\label{eq:def2point}
\end{eqnarray}
where FT denotes the spatial Fourier transform.
According to the FRG result, this function
takes the following form for large wavenumbers $k=|\vec k|$  
and small time delays:
\begin{eqnarray}
	C^{(2)}_S (t, \vec{k}) 
	&= C_S \epsilon^{2/3} k^{-11/3} \exp\Big\{-\alpha_S (L/\tau_0)^{2} t^2 k^2 \Big\} \quad
	\label{eq:theory2pcorrsmall}
\end{eqnarray}
and in the regime of large time delays:
\begin{equation}
C^{(2)}_L (t, \vec{k}) = C_L \epsilon^{2/3} k^{-11/3} \exp\left\{-\alpha_L (L^2/\tau_0) |t| k^2 \right\} 
\label{eq:theory2pcorrlarge}
\end{equation}
with \(\epsilon\) the energy dissipation rate, \(L\) the integral length scale,
$\tau_0 = (L^2/\epsilon)^{1/3}$  the eddy-turnover time at the integral scale,
and \(\alpha_{S,L}\) and $C_{S,L}$ nonuniversal constants -- the subscript $S$
and $L$ standing for `short time' and `long time' respectively. This expression
conveys that the velocity field decorrelates at small time delays as a Gaussian
of the variable \(t k\), whereas at large time delays, the decay of the
correlation function crosses over to an exponential in $t$.  As mentioned in
the introduction, the Gaussian behavior at small $t$ is
well-known from experimental data and numerical simulations and interpreted as
a consequence of the random sweeping effect. It turns out that the exponential
decay at large $t$  can also be simply understood in a similar framework, as
discussed in the Sec.~\ref{sec:heuristic}.
  
Let us comment on the domain of validity of these results. The factors in
curly brackets in Eqs.~\eqref{eq:theory2pcorrsmall} and
\eqref{eq:theory2pcorrlarge} are exact in the limit of large wavenumber $k\gg
L^{-1}$, which means that the corrections to these terms are at most of order
${\cal O}(k)$. We can
quantify more precisely where this limit is
reached using our DNS data. In contrast, the terms in front of the
exponential in Eqs.~\eqref{eq:theory2pcorrsmall} and
\eqref{eq:theory2pcorrlarge} are not exact
in these expressions, as
they can be corrected by higher-order contributions neglected in the large
wavenumber expansion. Otherwise stated, these expressions 
do not account for intermittency corrections on the
exponent 11/3, which merely corresponds to K41
scaling.

\par The FRG theory yields a  more general result: the spatio-temporal
dependence of any multi-point correlation functions of the turbulent velocity
field in the limit of large wavenumbers
\cite{tarpinBreakingScaleInvariance2018}. We concentrate in this work on the
three-point correlation function, defined as \begin{eqnarray}
	&C^{(3)}_{\alpha \beta \gamma}  (t_1, \vec{k}_1, t_2, \vec{k}_2)\equiv   \nonumber\\ 
	& \text{FT} \left[ \left<  u_\alpha(t_0 + t_1, \vec{r}_0 + \vec{r}_1) u_\beta(t_0 + t_2, \vec{r}_0 + \vec{r}_2) u_\gamma(t_0, \vec{r}_0) \right> \right]  \nonumber\\ 
	&= \left< \hat{u}_\alpha(t_0 + t_1, \vec{k}_1) \hat{u}_\beta(t_0 + t_2, \vec{k}_2) \hat{u}^*_\gamma(t_0, \vec{k}_1+\vec{k}_2) \right> 
	\label{eq:def3point}
\end{eqnarray}
where translational invariances in space and time follow
from the assumptions of homogeneity and stationarity.
In the limit where all the wavenumbers $k_1$, $k_2$, $|\vec{k}_1+ \vec{k}_2|$  are large,  the FRG calculation leads to the following form at small time delays $t_1$ and $t_2$
\begin{eqnarray}
	&C^{(3)}_{\alpha \beta \gamma}  (t_1, \vec{k}_1, t_2, \vec{k}_2) =\nonumber\\
	& C^{(3)}_{\alpha \beta \gamma}  (0, \vec{k}_1, 0, \vec{k}_2) \exp\left\{- \alpha_S (L/\tau_0)^{2}  \left|\vec{k}_1 t_1 + \vec{k}_2 t_2 \right|^2 \right\}
	\label{eq:theory3pcorr}
\end{eqnarray}
with $\alpha_S$ the same constant as in Eq.~\eqref{eq:theory2pcorrsmall}. 
Note that a similar expression as Eq.~\eqref{eq:theory2pcorrlarge} 
is also available
for large time delays, but it is not considered here since it is out of reach
of our simulations. In this work, we 
consider the simplified case
\(t = t_1 = t_2\), thus aiming at testing the theoretical form \begin{equation}
 C^{(3)}_{\alpha \beta \gamma}  (t, \vec{k}_1, t, \vec{k}_2) \sim  \exp\left\{- \alpha_S (L/\tau_0)^{2} \left|\vec{k}_1 + \vec{k}_2 \right|^2 t^2 \right\}.
 \label{eq:theory3pcorrtt}
 \end{equation}
One hence expects to observe that the three-point correlation functions
at large
wavenumbers are also Gaussian functions
of a variable \(|\vec{k}_1 + \vec{k}_2| t\) 
for small time delays $t$, with the same prefactor $\alpha_S$ as in
the two-point correlation functions.

Let us emphasize that similar results hold for any $n$-point correlation
functions at large wavenumbers, and are valid for arbitrary time
regimes, although for intermediate times the expressions take a more
complicated integral form \cite{tarpinBreakingScaleInvariance2018}.  Their
status is generically the same as discussed above for the two-point
correlations: 
The leading terms in the exponentials are exact in the limit of large
wavenumbers, whereas the prefactors
of these exponentials are not. 
Let us now give a simple physical interpretation of these
results.

\subsection{Physical interpretation}
\label{sec:heuristic}
The short-time predictions for time-dependence of two-point velocity correlations \eqref{eq:theory2pcorrsmall} and of three-point correlations 
\eqref{eq:theory3pcorr} were both given in an early analysis of Eulerian sweeping effects by Kraichnan 
\cite{kraichnanKolmogorovHypothesesEulerian1964}. As we show now, the novel prediction of  long-time exponential 
decay \eqref{eq:theory2pcorrlarge}
and similar long-time decay of general multi-point correlations were implicit in that earlier analysis,  
but unrecognized at the time. Both short-time and long-time decay regimes can be obtained from the following 
Lagrangian expression for the Eulerian velocity field
\begin{eqnarray}
 & u_i(t, \vec{r}) = \exp_{\to}[-\vec{\xi}(t,\vec{r}|t_0)\cdot\vec{\nabla}] u_i(t_0, \vec{r}) \nonumber \\ 
 & +\int_{t_0}^t ds\  \exp_{\to}[-\vec{\xi}(t,\vec{r}|s)\cdot\vec{\nabla}] \left[ \nu \nabla^2 u_i(s, \vec{r})-\nabla_ip(s,\vec{r})\right]. 
 \label{eq:convecteq}
 \end{eqnarray}
This is equation (7.7) in the paper of Kraichnan \cite{kraichnanKolmogorovHypothesesEulerian1964} when specialized to $s=t$ 
there (and with a minor typo corrected in the final term). Here $p(t,\vec{r})$ is the pressure, $\vec{\xi}(t,\vec{r}|s)=
\vec{r}-\vec{X}(t,\vec{r}|s)$ 
is the Lagrangian displacement vector, where 
 \begin{equation}
 \frac{d}{ds}\vec{X}(t,\vec{r}|s)=\vec{u}(s,\vec{X}(t,\vec{r}|s)), \quad \vec{X}(t,\vec{r}|t)=\vec{r}
 \end{equation}
 defines the position $\vec{X}(t,\vec{r}|s)$ at time $s$ of the Lagrangian fluid particle located at position $\vec{r}$
 at time $t.$ Finally $\exp_{\to}[-\vec{\xi}(t,\vec{r}|s)\cdot\vec{\nabla}]$ denotes an operator-ordered 
 exponential with all gradients $\vec{\nabla}$ ordered to the right and thus not acting upon the $\vec{r}$-dependence
 in $\vec{\xi}(t,\vec{r}|s)$. The intuitive meaning of equation \eqref{eq:convecteq} is that it ``states that the velocity field 
 at later times is the result of self-convection of the initial velocity field, together with convection of all of the 
 velocity increments induced at later times by viscous and pressure forces'' \cite{kraichnanKolmogorovHypothesesEulerian1964}.
 
 The formula \eqref{eq:convecteq} yields both of the FRG predictions \eqref{eq:theory2pcorrsmall} and \eqref{eq:theory2pcorrlarge}
 when some plausible statistical and dynamical assumptions are introduced. First, the displacement field $\vec{\xi}$ is expected
 to vary more slowly in space and time than the gradients of velocity $\vec{u}$ and of pressure $p$ that result from the 
 action of the exponential operator. Slowness in time allows one to factor out the exponential as 
 \begin{eqnarray}
 & u_i(t, \vec{r}) = \exp_{\to}[-\vec{\xi}(t,\vec{r}|t_0)\cdot\vec{\nabla}]\times \nonumber \\ 
 & \Big\{u_i(t_0, \vec{r}) +\int_{t_0}^t ds\ \left[ \nu \nabla^2 u_i(s, \vec{r})-\nabla_ip(s,\vec{r})\right]\Big\}. 
 \label{eq:convecteq2}
 \end{eqnarray}
 and slowness in space allows the Fourier transform to be evaluated as 
  \begin{eqnarray}
 & \hat{u}_i(t, \vec{k}) = \exp[-i\vec{\xi}(t,\vec{r}|t_0)\cdot\vec{k}]\times \nonumber \\ 
 & \Big\{\hat{u}_i(t_0, \vec{k}) -\int_{t_0}^t ds\ \left[ \nu k^2 \hat{u}_i(s, \vec{k})+k_i\hat{p}(s,\vec{k})\right]\Big\}. 
 \label{eq:convecteq3}
 \end{eqnarray}
 The next assumption is that the displacement field $\vec{\xi}$ is almost statistically independent of the 
 Fourier-transformed velocity fields at the initial time $t_0$, so that by the definition \eqref{eq:def2point}
 \begin{equation}
 C^{(2)}(t,\vec{k}) = \langle  \exp[-i\vec{\xi}(t,\vec{r}|0)\cdot\vec{k}]\rangle \Big\{C^{(2)}(0,\vec{k}) +{\cal O}(|t|)\Big\} .
  \end{equation}
 Finally, since the Lagrangian displacement is dominated by the largest scales of the turbulent flow, which 
 have nearly Gaussian statistics, it is plausible that $\vec{\xi}$ is also an approximately normal random field, so that  
 \begin{equation}
 C^{(2)}(t,\vec{k}) = \exp\left[-\frac{1}{2}\langle |\vec{\xi}(t,\vec{r}|0)|^2\rangle k^2 \right] \Big\{C^{(2)}(0,\vec{k}) +{\cal O}(|t|)\Big\}.
 \label{eq:Kr2pcorr} 
  \end{equation}
 According to this argument, the 2-point velocity correlation undergoes a rapid decay in the time-difference 
 $t$ which arises from an average over rapid oscillations in the phases of Fourier modes due to sweeping,
 or ``convective dephasing''\cite{kraichnanKolmogorovHypothesesEulerian1964}.
 
 The variance of the Lagrangian displacement in the exponent
 of \eqref{eq:Kr2pcorr} was the subject of a classical study by Taylor \cite{taylor1922diffusion} on 1-particle turbulent 
 dispersion. Exploiting the expression 
 \begin{equation}
 \vec{\xi}(t,\vec{r}|0)=\int_{0}^t \vec{u}(t,\vec{r}|s)\ ds, 
 \end{equation} 
two regimes were found:  
\begin{equation}
\langle |\vec{\xi}(t,\vec{r}|0)|^2\rangle \sim \left\{ \begin{array}{ll}
                                                                                 u_{RMS}^2  t^2 &  |t|\ll \tau_0 \cr
                                                                                 2D|t| &   |t|\gg \tau_0 
                                                                                 \end{array} \right. 
\label{xivar} \end{equation}
where the early-time regime corresponds to ballistic motion with the rms velocity $u_{RMS}$ and 
the long-time regime corresponds to diffusion with a turbulent diffusivity $D\propto u_{RMS}^2\tau_0.$
Using the relation $u_{RMS}\propto L/\tau_0$ and the result \eqref{eq:Kr2pcorr} for the 2-point velocity 
correlation, these two regimes of 1-particle turbulent dispersion correspond exactly to the short-time 
scaling \eqref{eq:theory2pcorrsmall} and the long-time scaling \eqref{eq:theory2pcorrlarge} predicted by FRG, with $\frac{1}{2}u_{RMS}^2=\alpha_S(L/\tau_0)^2$ and $D=\alpha_L L^2/\tau_0.$ To make more precise contact with the FRG analysis, one can 
introduce the temporal Fourier transform 
\begin{equation} 
\vec{v}(t,\vec{r};\omega)=\int ds\ e^{i\omega s} \vec{u}(t,\vec{r}|s)
\end{equation} 
and the corresponding (Lagrangian) frequency spectrum
$\langle \vec{v}(\omega)\cdot\vec{v}(\omega')\rangle=E(\omega)
\delta(\omega+\omega').$ It is then easy to see that
the displacement variance in \eqref{xivar} can be written as 
\begin{equation}
\langle |\vec{\xi}(t,\vec{r}|0)|^2\rangle
=\frac{1}{\pi}\int d\omega \ \frac{1-\cos(\omega t)}{\omega^2}
E(\omega) \label{int1} \end{equation}
in terms of the velocity spectrum. This formula should be compared 
with the leading-order FRG flow equation (30) of Tarpin et al. \cite{tarpinBreakingScaleInvariance2018} obtained in the limit 
of large wavenumber $|\vec{k}|$ 
\begin{equation} \kappa\partial_\kappa \ln C_\kappa^{(2)}(t,\vec{k}) 
=\frac{2}{3}|\vec{k}|^2
\int d\omega \ \frac{1-\cos(\omega t)}{\omega^2} J_\kappa(\omega) 
\label{int2} \end{equation} 
where the common factor inside the two frequency integrals yields identical 
short-time and long-time power-law asymptotics ($\propto t^2$ and $\propto t$, resp.) in both expressions \eqref{int1} and \eqref{int2}. 

The above arguments can obviously be applied to general multi-point velocity correlations, 
yielding similar results. They provide an intuitive physical interpretation of the two scaling regimes 
of the FRG results \cite{tarpinBreakingScaleInvariance2018}, with time-decay corresponding to a 
convective dephasing mechanism. In particular, the long-time exponential decay is suggested to
arise from the diffusive linear growth in the position variance of a Lagrangian particle advected by 
homogeneous turbulence. This long-time exponential decay regime appears to be a novel prediction 
of the FRG approach. For example,  it is quite distinct from the instantaneous exponential decay of the 2-point velocity 
correlator predicted by Rayleigh-Ritz analysis with a $K$-$\epsilon$ closure \cite{eyink1998evaluation},
which occurs on very short time-scale before convective dephasing can act and which is interpreted 
as an eddy-viscosity effect. Needless to say, the FRG derivation of  \eqref{eq:theory2pcorrsmall} and 
\eqref{eq:theory2pcorrlarge} is considerably more systematic and controlled than the heuristic 
argument presented in this section.

\section{Results of direct numerical simulations}
\label{sec:DNS}

We perform direct numerical simulations (DNS) of a stationary 3D
incompressible homogeneous and isotropic turbulent flow. The computation domain
represents a cube of size \(2\pi\) with periodic boundary conditions. 
We use five
values of the Taylor-scale Reynolds number: 
\(R_\lambda =  40, 60,
90, 160, 250\) with corresponding spatial grid size \(N^3 = 64^3, 128^3, 256^3,
512^3, 1024^3\) (see Table~\ref{tab:params}). The spatial resolution of all
simulations fulfills the condition \(k_{max} \eta \simeq 1.5\), where \(k_{max}
= N/2\) is the maximal wavenumber in the simulation and \(\eta\) is the
Kolmogorov length scale. The incompressible Navier-Stokes equation is solved
numerically with the use of a pseudospectral method in space
\cite{canutoSpectralMethodsEvolution2007} and a second order Runge-Kutta scheme
of time advancement. To achieve a statistically stationary state, the velocity
field is randomly forced at large scales
\cite{alveliusRandomForcingThreedimensional1999}. 
We perform a de-aliasing with the use of the polyhedral truncation method
\cite{canutoSpectralMethodsEvolution2007}.  
\begin{table}
	\centering
	\begin{ruledtabular}
	\begin{tabular}{ccccccccc}
		\(R_{\lambda}\) & N & \(\nu\) & \(u_{RMS}\) & \(\tau_0\) & \(\Delta t\) & \(\Delta T_w\) & \(N_t\) & \(K_c L\) \\
		\hline
		40 & 64 & \(10^{-4}\) &0.0059 & 245 & 0.9 & 400 & 1008& 14.6 \\
		60 & 128 & \(10^{-4}\) &0.0147 & 134 & 0.1 & 75 & 665 & 23.7 \\
		90 & 256 & \(10^{-4}\) &0.0375 & 45.3 & 0.03 & 10.0 & 624 & 42.5 \\
		160 & 512 & \(10^{-4}\) &0.0974 & 19.0 & 0.005 & 1.0 & 322 & 74.4 \\
		250 & 1024 & \(10^{-4}\) &0.2482 & 7.24 & 0.001 & 0.2 & 33 & 144 \\
	\end{tabular}
	\end{ruledtabular}
	\caption{Parameters of simulations for the analysis of two-point and
three-point correlations at small time delays. \(R_\lambda\) - Taylor-scale
Reynolds number, \(N\) - spatial grid resolution, \(\nu\) - kinematic
viscosity, \(u_{RMS}\) - root mean square velocity, \(\tau_0\) - eddy turnover
time at the integral scale, \(\Delta t\) - simulation time step, 
\(\Delta T_w \) - width of a time window of correlation observation, \(N_t\) 
- number of
recorded time windows, \(K_c L\)- nondimensional  cut-off wavenumber of the
scale decomposition.} 
\label{tab:params}
\end{table}

\subsection{Two-point spatio-temporal correlations at small time delays}
\label{sec:2point}
Once the simulations reach a statistically steady state, we 
compute the velocity correlation functions with the following method: 
At a chosen time \(t_0\)
we store the spectral 3D vector velocity field in the memory.
At the next iterations the updated velocity field at 
time \(t_0 + i\Delta t\) 
is multiplied point-wise by the velocity field at time~\(t_0\). 
Since the velocity field is statistically
isotropic,
the two-point velocity correlation function is computed by averaging
over spherical spectral shells \(S_n\) of thickness
   \(\Delta k = 1\) so that \(\vec{k} \in S_n \ \text{if} \ n-1 <
\big|\vec{k}\big|< n, n = 1, .., N/2\). After a certain number of time
iterations, when the magnitude of the correlations at all scales of interest
is close to zero, the reference time \(t_0\) 
is redefined as the current time,
and the reference velocity field in the memory is updated. The resulting
correlation function is averaged over time windows with different reference
times~\(t_0\), and the real part is taken: 
\begin{equation}
	\bar{C}^{(2)} (t,k) = \frac{1}{N_t} \sum_{j=1}^{N_t} \frac{1}{M_n}  \sum_{\vec{k} \in S_n} {\mathrm{Re}} \left[ \hat{u}_i({t_0}_j, \vec{k}) \hat{u}^*_i({t_0}_j + t, \vec{k}) \right]
	\label{eq:def2pcorrNumerical}
\end{equation}
where \(N_t\) is the number of time windows in the simulation, \(M_n\) is the number of modes in the  spectral spherical shell \(S_n\), and \(k = n\Delta k, n \in \mathbb{Z}\) .
We hence  obtain a numerical estimation of the two-point spatio-temporal correlation function \(C^{(2)}\) defined in Eq.~\eqref{eq:def2point} 
with averaging in space and time.
\begin{figure}
	\centering
	\includegraphics[width=0.8\linewidth]{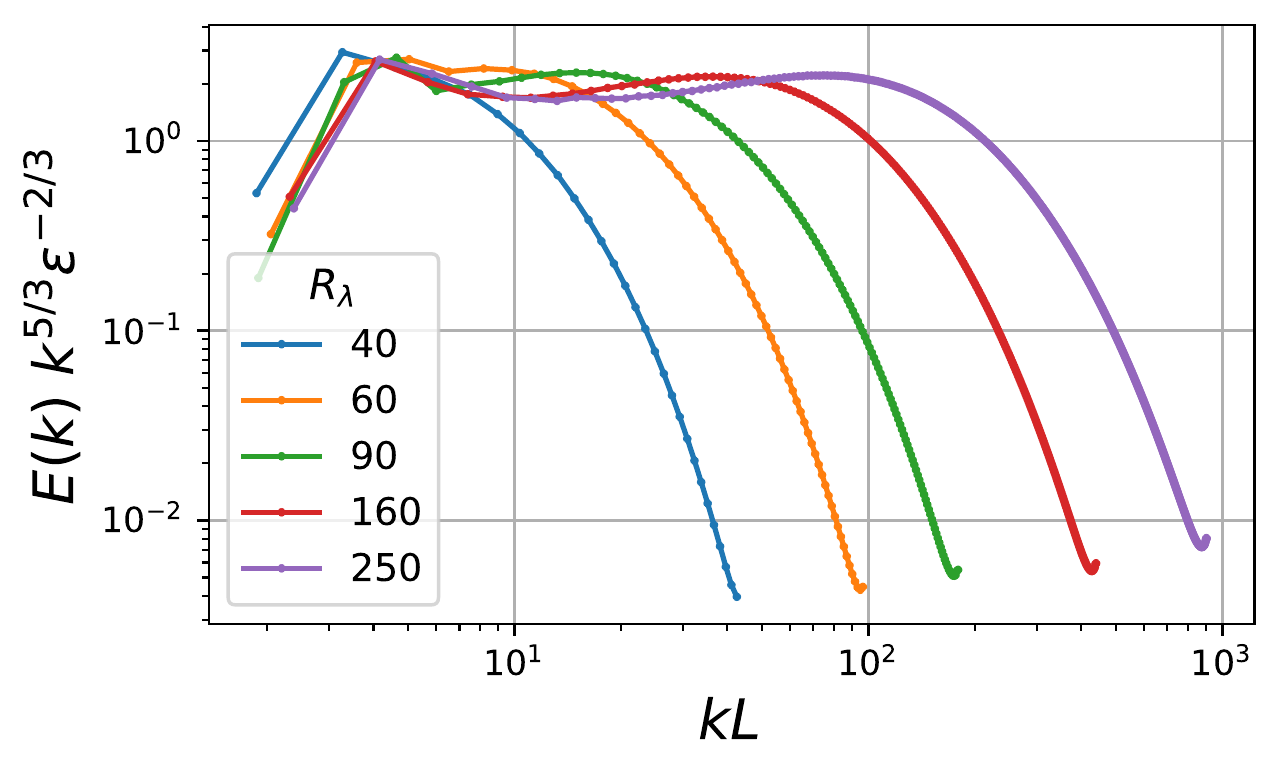}
	\caption{Compensated spatial spectrum of the kinetic energy obtained
from the averaged two-point spatio-temporal correlation function \(C^{(2)}\) at
zero time delay according to Eq.~\eqref{eq:spectrum}. \(\epsilon\) is the
energy dissipation rate, \(L\) the integral length scale, and $R_\lambda$  the
Reynolds number at the Taylor microscale.} 
\label{fig:spectra}
\end{figure}
\par Note that at \(t=0\), the integration over a spherical shell in spectral space of the correlation function \(C^{(2)}\) in Eq.~\eqref{eq:def2point} 
gives the spectrum of kinetic energy:
\begin{equation}
	E(k) = 4 \pi k^2 C^{(2)}(t=0, k) = 4 \pi C_S \epsilon^{2/3} k^{-5/3}
	\label{eq:spectrum}
\end{equation}
The compensated spatial spectra obtained from the averaged two-point 
spatio-temporal correlation function at zero time delay are shown in 
the Fig.~\ref{fig:spectra}. 
The inertial regimes of these spectra approximately conform to 
the Kolmogorov 5/3 power-law decay and are followed by the
dissipation regime. While there is no visible inertial
range at the lowest $R_\lambda$, 
it extends over about one decade at the largest $R_\lambda$.
We first focus on the behavior of the correlation function
\(C^{(2)}\) at small time delay, and we normalize all data by the correlation
function for coincident times
\(C^{(2)}(t=0,k)\).  \begin{figure}
	\centering
	\includegraphics[width=0.75\linewidth]{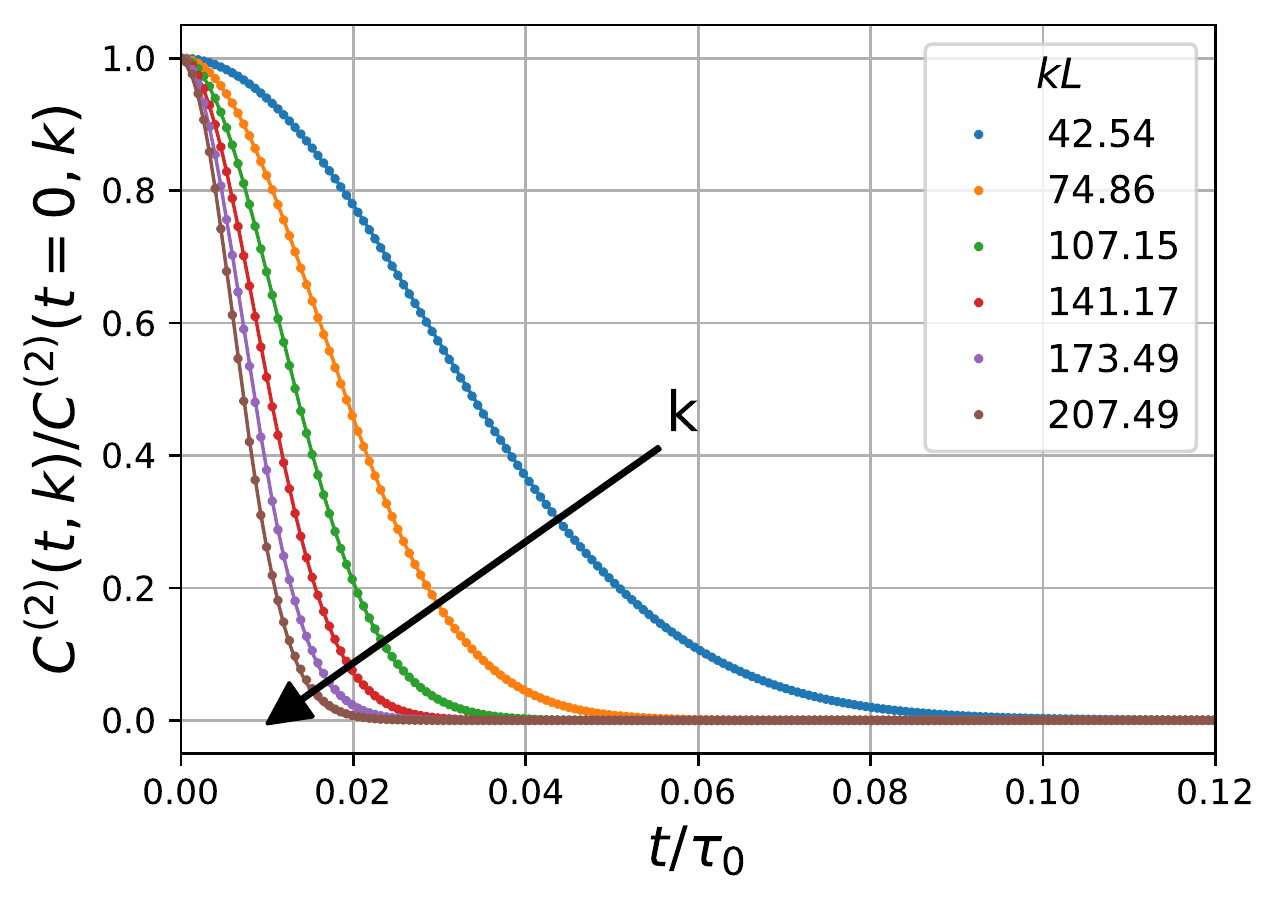}\\
	\includegraphics[width=0.75\linewidth]{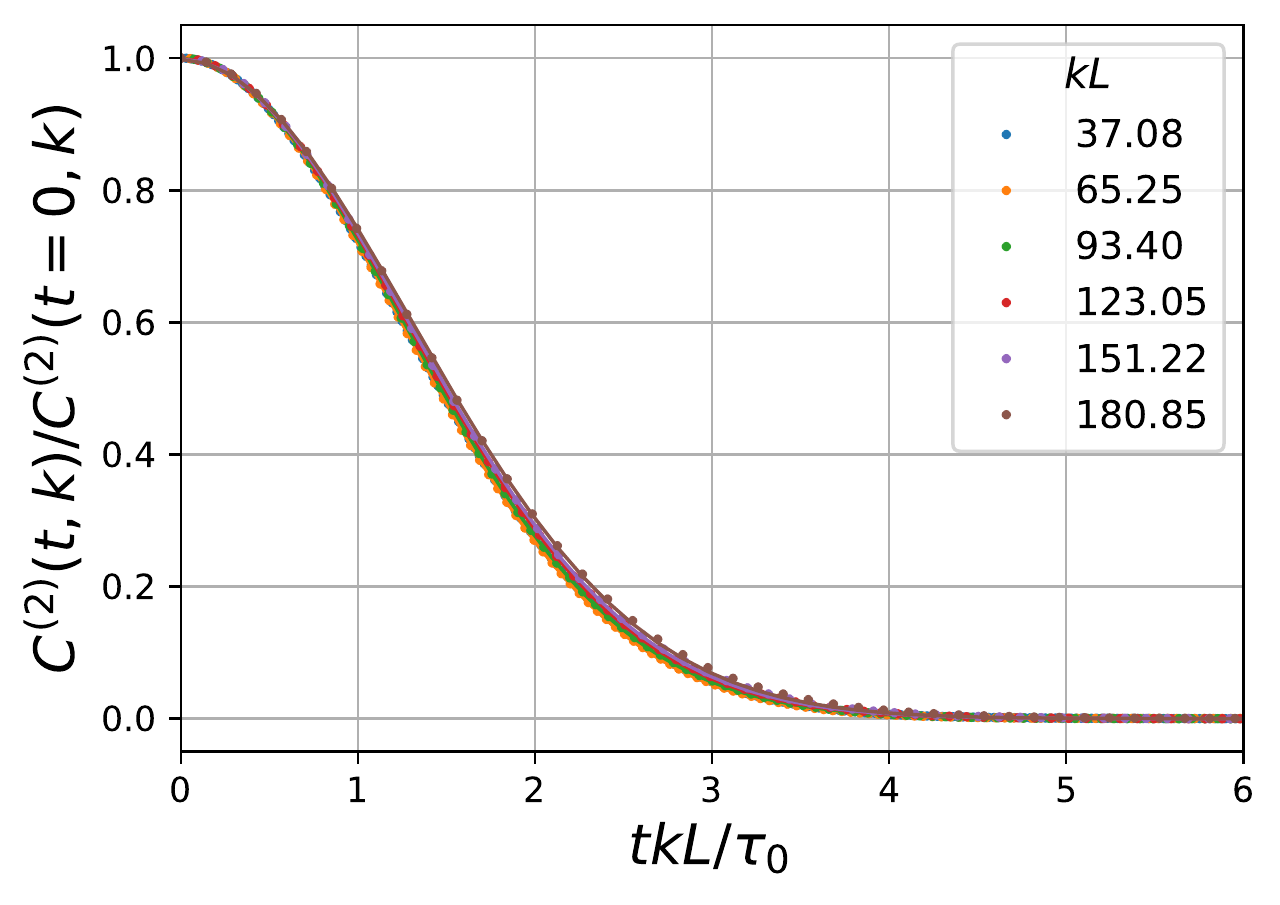}\\
	\caption{
Time dependence of the normalized two-point correlation function \(C^{(2)} (t,k)\) 
	at different wavenumbers \(k\) in the simulation at \(R_\lambda = 90\). Upper panel:
	data from the numerical simulation denoted with dots and its Gaussian fits denoted
	with continuous lines; bottom panel: the same data plotted as a function of the scaling variable \(tk\),
	which results in the collapse of all the curves, as expected from 
Eq.~\eqref{eq:theory2pcorrsmall}. 
	\(L\) is the integral length scale, \(\tau_0\) is the large eddy-turnover time scale.}
	\label{fig:2pcorr}
\end{figure}
\par According to the theoretical expression Eq.~\eqref{eq:theory2pcorrsmall}, 
we expect a Gaussian dependence in $t$
 at small time delays, which we precisely observe in all our simulations. 
We show in the Fig.~\ref{fig:2pcorr}
an example of the numerical results for $C^{(2)}$ at various wavenumbers for \(R_\lambda = 90\).
All curves display a Gaussian behavior, the fits are analyzed in details below.
Prior to this, let us comment on the  scaling. When plotted as a function of
the variable $tk$, all the curves collapse onto a single Gaussian, as expected
from the  Eq.~\eqref{eq:theory2pcorrsmall}. This is illustrated in the
bottom panel of Fig.~\ref{fig:2pcorr}. We emphasize that this $tk$ scaling of
the correlation function is different from the \(tk^{2/3}\) scaling that
one would obtain from dimensional considerations based on the
standard assumptions of Kolmogorov's theory of turbulence, taking as the only
relevant parameters the energy dissipation rate \(\epsilon\) and the wavenumber 
\(k\).
\footnote{In fact, Kolmogorov in his original 1941 paper 
\cite{kolmogorov1941local} emphasized that such dimensional 
reasoning should apply to multi-time correlations only in a quasi-Lagrangian frame.}  
As explained in Sec.~\ref{sec:heuristic}, the \(tk\) scaling arises from dimensional
analysis if the root-mean-square velocity \(u_{RMS} \sim L/\tau_0\) is also
included as a relevant parameter. This constitutes an explicit
breaking of scale invariance, which originates in the random sweeping.
\(u_{RMS}\) is indeed the characteristic velocity scale of the random advection
process of small-scale velocities by large vortices. 
 
\par We now turn to the analysis of the Gaussian fits. The time correlation
curves at various wavenumbers and various Reynolds numbers are fitted 
using the nonlinear least-square method (Levenberg–Marquardt algorithm), with
the Gaussian fitting function: {\(f_{s} (t) = c e^{-(t/\tau_s)^2}\)} 
where~$\tau_s$ and~$c$ are the fitting parameters. 
Performing a nondimensionalization with parameter
\(L/\tau_0 \approx u_{RMS}\) renders the correlation function plots at
various Reynolds number comparable. The fitting range for all the data sets
corresponds to the range of nondimensional variable \((t k L/\tau_0) \in [0,
2.5]\), within this range, all the correlation functions are
accurately modelled by the Gaussian $f_s$.

\begin{figure}
	\centering
	\includegraphics[width=0.75\linewidth]{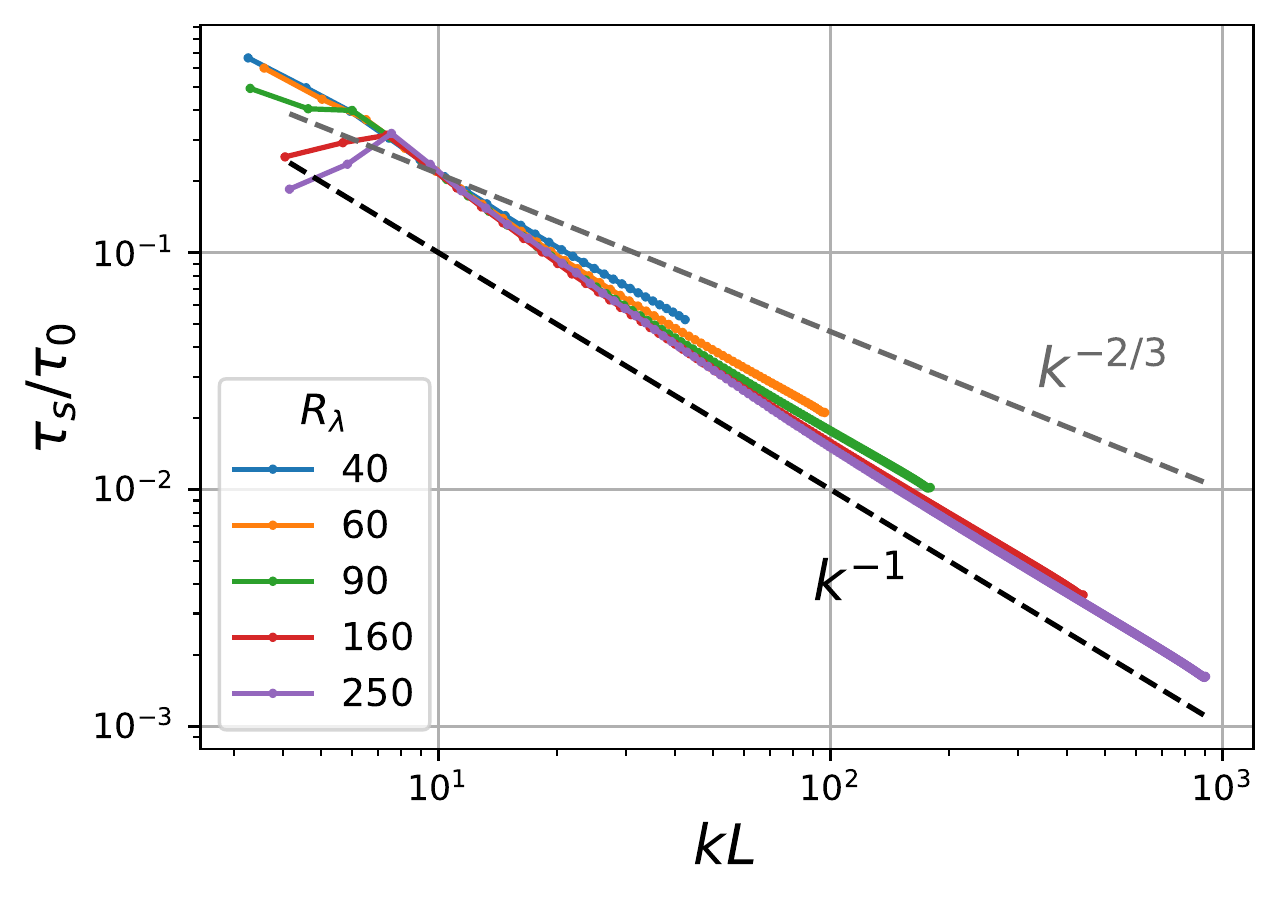}
	\caption{ Dependence of the decorrelation time $\tau_s$ resulting from the Gaussian fit on the wavenumber
	 in log-log scale for various $R_\lambda$. Times on the vertical axis are normalized by the large eddy 
	 turnover time scale $\tau_0$, the wavenumber on the horizontal axis is normalized by the integral length scale $L$. }
	\label{fig:2pcorrdecorrtime}
\end{figure}
The fitting parameter \(\tau_s\) is the characteristic time scale of
the correlation function, its dependence on the wavenumber \(k\) is shown in
Fig.~\ref{fig:2pcorrdecorrtime} for various $R_\lambda$.
While for
small wavenumbers the dependence is not regular, at intermediate and large
wavenumbers the decorrelation time clearly decays as \(k^{-1}\). This result
confirms that the collapse in the Fig.~\ref{fig:2pcorr} occurs for the
\(tk\)-scaling. It is also in plain agreement with a similar analysis performed
in Ref. \onlinecite{favierSpaceTimeCorrelations2010}. 

\begin{figure}
	\includegraphics[width=0.75\linewidth]{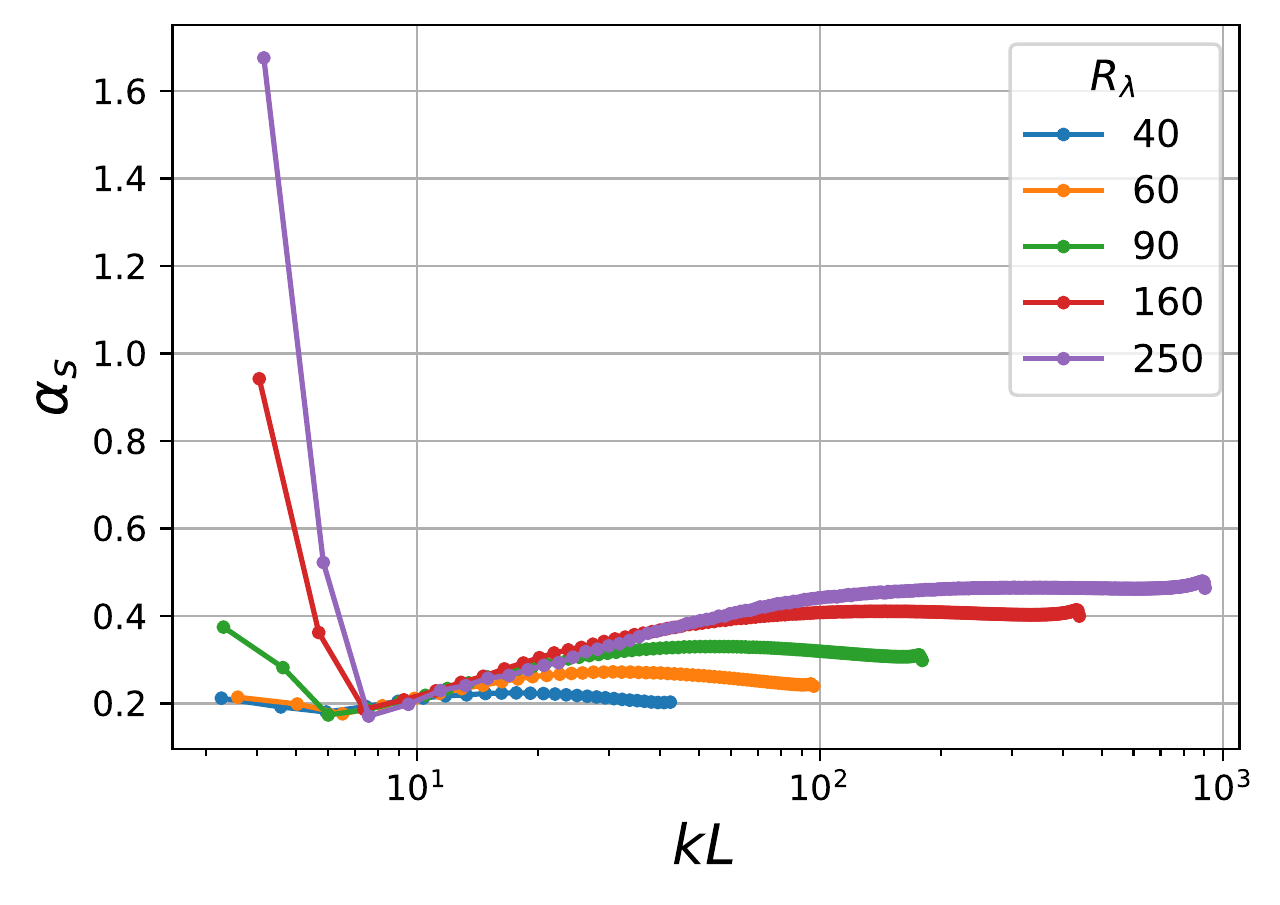}
	\caption{Estimation of the theoretical parameter \(\alpha_S\) in the Eq. (\ref{eq:theory2pcorrsmall})
	 from the results of the fit of numerical data. }
	\label{fig:alphasvsk}
\end{figure}
\begin{figure}
	\includegraphics[width=0.75\linewidth]{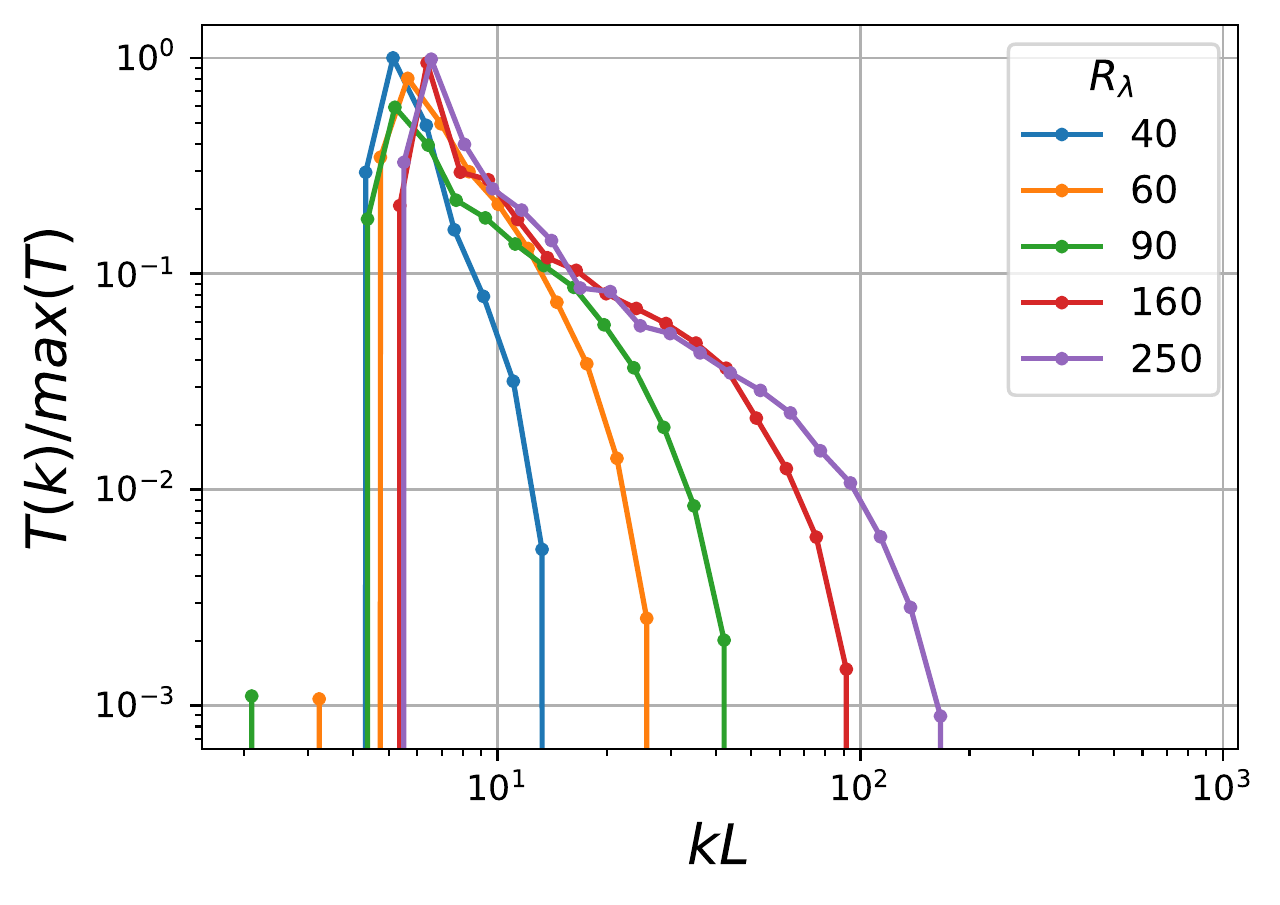}
	\caption{ Rate of the direct energy transfer from the forcing range to the wavenumber $k$ normalized by 
	the maximal value at various Reynolds numbers computed with the use of the shell-to-shell energy transfer
	 method described in Ref. \onlinecite{vermaEnergyTransfersFluid2019}.}
	\label{fig:forcingTransfer}
\end{figure}

\par One can estimate the coefficient \(\alpha_S\) in the theoretical
expression Eq.~\eqref{eq:theory2pcorrsmall} from the fits:
{\(\alpha_S \approx (\tau_0/\tau_skL)^2\)}. Plotting \(\alpha_S\)
versus \(k\) as in Fig.~\ref{fig:alphasvsk} shows that the numerical estimation
of \(\alpha_S\) reaches a plateau  at large wavenumbers, which length
increases with $R_\lambda$. 
\par Whereas the Gaussian regime can be observed already at intermediate wavenumbers, the
value of $kL$ at which $\alpha_S$ settles 
 to this plateau appears to be dependent on $R_\lambda$.
 The deflection of $\alpha_S$ from the plateau value at the intermediate wavenumbers can be attributed to the effect of
 the forcing in the numerical scheme. This can be observed from the analysis of the direct energy transfers with the modes of the forcing range, as shown on  
the Fig.~\ref{fig:forcingTransfer}. The "ideal" numerical simulation would exhibit
a single peak of energy transfers close to the forcing range itself, indicating
the presence of
the local modal interactions only, when the smaller scales receive energy only
through the turbulent energy cascade. However, 
Fig.~\ref{fig:forcingTransfer}
shows that direct energy transfers occur
not only in the closest vicinity of the forcing range, 
but also at a significant level over a band
of wavenumbers, the width of which
depends on $R_\lambda$.
This means that the wavenumbers from this band are subjected not only to the
local energy cascade, but also to nonlocal direct energy transfers from the
forcing range. The occurrence of the nonlocal energy transfers in DNS can be a
consequence of the velocity forcing concentrated in a narrow spectral band at
large scales, as discussed in Ref.
\onlinecite{kuczajNonlocalModulationEnergy2006}.
  
\par This additional non-local interaction process slows down the velocity
decorrelation and results in lower values of $\alpha_S$. Matching the
horizontal axes of the figures \ref{fig:alphasvsk} and \ref{fig:forcingTransfer}
shows that the parameter $\alpha_S$ reaches a constant value at wavenumbers
where the direct energy exchanges with the forcing modes become 
negligible. We
can draw from these observations that the `large wavenumber' regime of the
theory can be here identified as the values of  $kL$ such that direct energy
transfers from the forcing range are negligible.

\par Let us summarize this part. The
data obtained from the DNS accurately confirm the theoretical 
expression \eqref{eq:theory2pcorrsmall}
for the two-point spatio-temporal correlations of the turbulent velocity field
for various scales at small time delays. In particular, the numerical data
show that the theoretical parameter \(\alpha_S\) reaches 
a plateau at large wavenumbers, in agreement with the theoretical  result. 
\subsection{Three-point spatio-temporal correlations at small time delays}
\label{sec:3point}
\par In this part, we estimate the three-point spatio-temporal correlations 
$C^{(3)}$ of the turbulent velocity field from the DNS data. 
The definition of \(C^{(3)}\) involves a product of Fourier transforms of the velocity field 
\(\hat{u}(t,\vec{k})\) at three different wavevectors: 
\begin{eqnarray}
	& C_{\alpha \beta \gamma}^{(3)}(t_1, \vec{k}_1, t_2,\vec{k}_2)  \equiv \nonumber \\ 
	& \left< \hat{u}_\alpha(t_0 + t_1, \vec{k}_1) \hat{u}_\beta(t_0 + t_2, \vec{k}_2) \hat{u}^*_\gamma(t_0, \vec{k}_1+\vec{k}_2) \right> 
	\label{eq:3pcorr_def}
\end{eqnarray}
In contrast with the two-point correlation function \(C^{(2)}\), 
the product in the expression \eqref{eq:3pcorr_def} is not local in
\(\vec{k}\). When parallel computation and parallel memory distribution are
used, the access to nonlocal quantities requires the implementation of
additional communication operations between the processors during the
simulation. This  implies a great increase of computation time and memory. In
order to avoid these additional implementation difficulties and computational
costs, we study and exploit a local three-point velocity quantity
naturally arising from the Navier-Stokes equation and already introduced in
earlier works \cite{kraichnanStructureIsotropicTurbulence1959}.
\subsubsection*{Advection-velocity correlation function}
\par The Navier-Stokes equation in the spectral space can be written as:
\begin{equation}
	\partial_t \hat{u}_\ell (t, \vec{k}) = \hat{N}_\ell (t, \vec{k}) - \nu k^2  \hat{u}_\ell (t, \vec{k}) + \hat{f}_\ell  (t, \vec{k})
	\label{eq:ns_spec}
\end{equation}
where \(\hat{N}_\ell (t, \vec{k}) = -i k_n P_{\ell m} \sum_{k^\prime} \hat{u}_m (t, \vec{k}^\prime) \hat{u}_n(t, \vec{k}-\vec{k}^\prime)\) is the Fourier transform of the advection and pressure gradient terms of the Navier-Stokes equation, \(P_{ij} = \delta_{ij} - k_i k_j/k^2\) is the projection tensor and \(\hat{f}_\ell\) is the spectral forcing. 
Multiplying Eq.~\eqref{eq:ns_spec} by
the conjugated velocity \(\hat{u}^*_\ell(t_0, \vec{k})\) at a fixed time
\(t_0\) and performing an ensemble average leads to the following equation for the two-point 
spatio-temporal correlation function \(C^{(2)} (t, \vec{k})\):
\begin{equation}
	(\partial_t + \nu k^2) C^{(2)} (t, \vec{k}) = \hat{T}(t, \vec{k}) + \hat{F} (t, \vec{k})
	\label{eq:c2}
\end{equation}
where \(\hat{T}(t, \vec{k}) \equiv \left<\hat{N}_i(t + t_0, \vec{k})
\hat{u}^*_i(t_0, \vec{k}) \right>\) is the spatio-temporal correlation of the
advection term and velocity, and \(\hat{F} (t, \vec{k}) = \left<\hat{f}_i(t +
t_0, \vec{k}) \hat{u}^*_i(t_0, \vec{k}) \right>\) is the spatio-temporal
correlation of the spectral forcing and velocity. Note that if the time delay
is set to zero (\(t = 0\)), then Eq.~\eqref{eq:c2} simplifies to the equation of
evolution of the average kinetic energy of a single spectral mode \(E_{kin}
(\vec{k}) = \frac{1}{2} C^{(2)} (0, \vec{k})\). This energy splits into
\(\frac{1}{2}\hat{T}(0,\vec{k})\) (the average nonlinear energy transfer
between modes) and \(\frac{1}{2}\hat{F} (0,\vec{k})\) (the average forcing
power input, which is assumed to be zero beyond the forcing range at large
scales).  

\par The advection-velocity correlation function \(\hat{T}\) is a
three-point statistical quantity, and its link with the three-point correlation
function \(C^{(3)}\) becomes clear if one develops the nonlinear term in the
definition of \(\hat{T}(t,\vec{k})\):
\begin{eqnarray}
	&\hat{T}(t, \vec{k}) \equiv \left<\hat{N}_\ell(t_0 + t, \vec{k}) \hat{u}^*_\ell(t_0, \vec{k}) \right> \nonumber \\
	&= -i k_n P_{\ell m} \sum_{k^\prime}  \left< \hat{u}_m (t + t_0, \vec{k}^\prime) \hat{u}_n(t + t_0, \vec{k}-\vec{k}^\prime) \hat{u}^*_\ell(t_0, \vec{k}) \right> \nonumber\\ 
	&= -i k_n P_{\ell m} \sum_{k^\prime} C^{(3)}_{m n \ell} (t, \vec{k^\prime}, t, \vec{k} - \vec{k}^\prime)
	\label{eq:link_c3_t}
\end{eqnarray} 
Hence, the correlation function $\hat{T}$ actually provides a linear
combination of three-point correlation functions. 
The theoretical prediction~\eqref{eq:theory3pcorrtt}
suggests that this type of sum of $C^{(3)}$ must be Gaussian 
(at small times and large wavenumbers). 
Thus, if the theoretical prediction is valid, 
one would expect that the appropriately computed correlation 
function $\hat{T}$ is also a Gaussian of the variable $tk$: 
\begin{equation}
    \hat{T}(t, \vec{k}) \sim  \sum_{k^\prime} C^{(3)}_{m n \ell} (t, \vec{k^\prime}, t, \vec{k} - \vec{k}^\prime)  \sim  \exp\left\{- \alpha_S (L/\tau_0)^{2} |\vec{k}|^2 t^2 \right\}
\end{equation}

\par Another useful property of the correlation function $\hat{T}$ is its link with the two-point correlation function $C^{(2)}$. Considering a small time delay \(t\), one can use the expression of the two-point correlation function \(C_s^{(2)}(t, \vec{k})\) of Eq.~\eqref{eq:theory2pcorrsmall}. 
Inserting this result into Eq.~\eqref{eq:c2} leads to an explicit expression 
for the function \(\hat{T}\) at small time delays (and for wavenumbers 
outside the forcing range):
\begin{eqnarray}
	&\hat{T}(t, \vec{k}) = \nu k^2 C^{(2)}(0, \vec{k}) \left( 1 - \frac{2 \alpha_S L^2}{\tau_0^2 \nu} t\right) \exp\Big\{-\alpha_S (L/\tau_0)^2 k^2 t^2\Big\} = \nonumber \\
	& = \hat{D}(\vec{k}) \left( 1 - 2 \alpha_S Re \cfrac{t}{\tau_0} \right) \exp\Big\{-\alpha_S (L/\tau_0)^2 k^2 t^2\Big\} 
	\label{eq:tnonfiltered}
\end{eqnarray}
where $\hat{D}(\vec{k}) = \nu k^2 C^{(2)}(0, \vec{k})$ is the spectral dissipation rate and 
$Re = \frac{U_{RMS} L}{\nu}$ is the Reynolds number. 
Eq.~\eqref{eq:tnonfiltered} indicates that the function $\hat{T}$ is 
in general not symmetric with respect to the origin of the $t$-axis, 
and that it can have a minimum and maximum at non-zero time delay $t$. 
 
\par To sum up, the advection-velocity correlation function \(\hat{T}\) is a
local quantity in spectral space, as it implies the multiplication of the
advection and velocity fields at the same  wave vector \(\vec{k}\), and it is
related to a sum of three-point nonlocal velocity correlation functions. The
equivalence of the function $\hat{T}$ at zero time delay to the spectral
energy transfer function and its link with the two-point spatio-temporal
correlation function Eq.~\eqref{eq:c2} facilitate the testing of the numerical
method and the interpretation of the results in the following. Note that an
equation similar to Eq.~\eqref{eq:c2} is also used in the Direct Interaction
Approximation scheme (DIA)  \cite{kraichnanStructureIsotropicTurbulence1959},
where a time dependent triple statistical moment similar to \(\hat{T}\) is
introduced.

\subsubsection*{Numerical method}
\par In the numerical simulations, 
we compute the correlation function \(\hat{T}(t,\vec{k})\) 
by point-wise multiplication of the Fourier transform of the nonlinear 
term \(\hat{N}(t_0 + t, \vec{k})\) by the velocity field \(\hat{u}^*(t_0, \vec{k})\). 
This quantity is local in spectral space and the computation does not require 
significantly more computational resources. 

\par 
We use the method already described in the Sec.~\ref{sec:2point}
to collect and average the data. However, note that in this case
it becomes necessary to take into account the sign of the time delay.
The advection-velocity correlation function
\(\hat{T}\) at negative time delays can be computed just by switching the time
instants of the fields in the following way: 
\begin{equation} 
\hat{T}(t, \vec{k}) =
\begin{cases}
\left<\hat{N}_i(t_0 + |t|, \vec{k}) \hat{u}_i^* (t_0, \vec{k}) \right>, &  t > 0 \\
\left<\hat{N}_i(t_0, \vec{k}) \hat{u}_i^* (t_0 + |t|, \vec{k}) \right> , & t < 0
\end{cases} 
\end{equation}
Hence, to compute
the correlation $\hat{T}(t, \vec{k})$ at negative time delays
during the simulation one only needs to store
the spectral advection field at one reference time~$t_0$. 

\subsubsection*{Scale decomposition}
\par Although the advection-velocity correlation function \(\hat{T}\) provides
a three-point statistical quantity that can be easily accessed in the numerical
simulations, it contains a summation  coming from the convolution in
the advection term  Eq.~\eqref{eq:link_c3_t}.  Contributions from all possible
wavevector triads \(\{\vec{k}^\prime, \vec{k}-\vec{k}^\prime, \vec{k}\}\) of
any scale are thus summed up. However, the FRG prediction 
Eq.~\eqref{eq:theory3pcorr} is valid in the limit where all three wavenumbers
are large. One hence needs to refine this sum in order to eliminate
contributions from the small wavenumbers.  

\par The simplest way to solve this issue is to
perform a scale decomposition of the velocity fields. We choose
a threshold wavenumber \(K_c\), so that all wavevectors of smaller norm
\(|\vec{k}|< K_c\) are considered as "large" scales and are denoted with a
superscript $L$, whereas the modes with higher wavenumbers are considered as
"small scales" and denoted with $S$. The velocity field is decomposed into
small- and large-scale parts $\vec{u} = \vec{u}^L + \vec{u}^S$. In the spectral
domain the decomposition is performed by a simple box-filtering operation: 
\begin{equation}
	\hat{u}^L_i (\vec{k}, t)= 
	\begin{cases}
		\hat{u}_i (\vec{k}, t), &   |\vec{k}| < K_{c} \\
		0, & |\vec{k}| \ge K_{c}
		\end{cases}
		\ \hat{u}^S_i (\vec{k}, t)= 
		\begin{cases}
		0, & |\vec{k}| < K_{c} \\
		\hat{u}_i (\vec{k}, t), &  |\vec{k}| \ge K_{c} \\
	\end{cases}
	\label{eq:udecompose}
\end{equation}
\par The velocity field scale decomposition leads to a decomposition of the advection-velocity correlation function $\hat{T}$ into four terms (here written as an example for a wavevector $\vec{k}$ belonging to the "small" scales):
\begin{equation}
	\hat{T}(\vec{k}, t) = \left[\hat{T}^{SSS} + \hat{T}^{SLS} + \hat{T}^{SSL} + \hat{T}^{SLL}\right](\vec{k},  t) 
	\label{eq:tdecompos}
\end{equation}
with $\hat{T}^{XYZ} (\vec{k}, t, t_0) = -{[\hat{u}_i^X]^*}(\vec{k},t_0) \ \text{FT}[u_j^Y \partial_j u_i^Z](\vec{k},t_0 + t)$ where $X,Y,Z$ stand for $S \ \text{or}\ L$. 

\par A similar decomposition at equal times has been used in studies of the
energy transfer function (Ref.
\onlinecite{frischTurbulenceLegacyKolmogorov1995,
vermaEnergyTransfersFluid2019}). Using the terminology of Ref.
\onlinecite{vermaEnergyTransfersFluid2019} for energy transfers, the first
superscript of a decomposition term is related to the mode receiving energy in
a triadic interaction process (it is actually the mode $\vec{k}$ for which the
equation \eqref{eq:tdecompos} is written setting $t=0$), the intermediate
superscript denotes the mediator mode and the last superscript is related to
the giver mode that sends the energy to the receiver mode. The mediator mode
does not loose nor receive energy in the interaction, it is related to the
velocity field which comes as prefactor of the operator nabla in the nonlinear
term of the Navier-Stokes equation, so one can term it the "advecting" field.

\par Let us give a 
physical interpretation of the terms of this decomposition. The term
$\hat{T}^{SSS}$ gathers all triadic interactions where the three modes
belong to the small scales. The term $\hat{T}^{SLS}$ is related to the energy
transfers between two small scales mediated by large scale modes. Both energy
transfers $\hat{T}^{SSS}$ and $\hat{T}^{SLS}$ occur between small scales, and
are thus supposed to be local in spectral space, so they form the turbulent
energy cascade. The terms $\hat{T}^{SSL}$ and $\hat{T}^{SLL}$
denote the direct energy transfers from large scale modes to small scale
modes, thus non-local interactions that we expect to be small
compared to the local interactions.
Let us now focus on the all-small scale term~$\hat{T}^{SSS}$,
which corresponds to the limit of large wavenumbers on which the theoretical
prediction relies.

\par The cut-off wavenumber $K_c$ of the filter in the
Eq.~\eqref{eq:udecompose}
is chosen in such a way that at \(k \gtrsim K_c\) the direct energy transfer
between small scale modes and those
of the forcing range (shown in the Fig.~\ref{fig:forcingTransfer})
becomes negligible. We expect that the dynamics of the modes at \(k \gtrsim
K_c\) does not depend directly on the forcing mechanism and we should observe
an approach to the universal behavior predicted by the theory. The value of
\(K_cL\) used for each simulation is provided in Table~\ref{tab:params}. 
The wavenumbers \(k \gtrsim K_c\) approximately correspond to the range of
validity of the theoretical prediction for the two-point correlation function
at large wavenumbers, as discussed in the Sec.~\ref{sec:2point}.

\subsubsection*{Results for the temporal correlations}
\par The data presented in this section are obtained from
the same set of simulations used for the analysis of the two-point correlation function at small time delays in  Sec.~\ref{sec:2point} and described in 
Table~\ref{tab:params}. 
\begin{figure}
	\centering
	\includegraphics[width=0.75\linewidth]{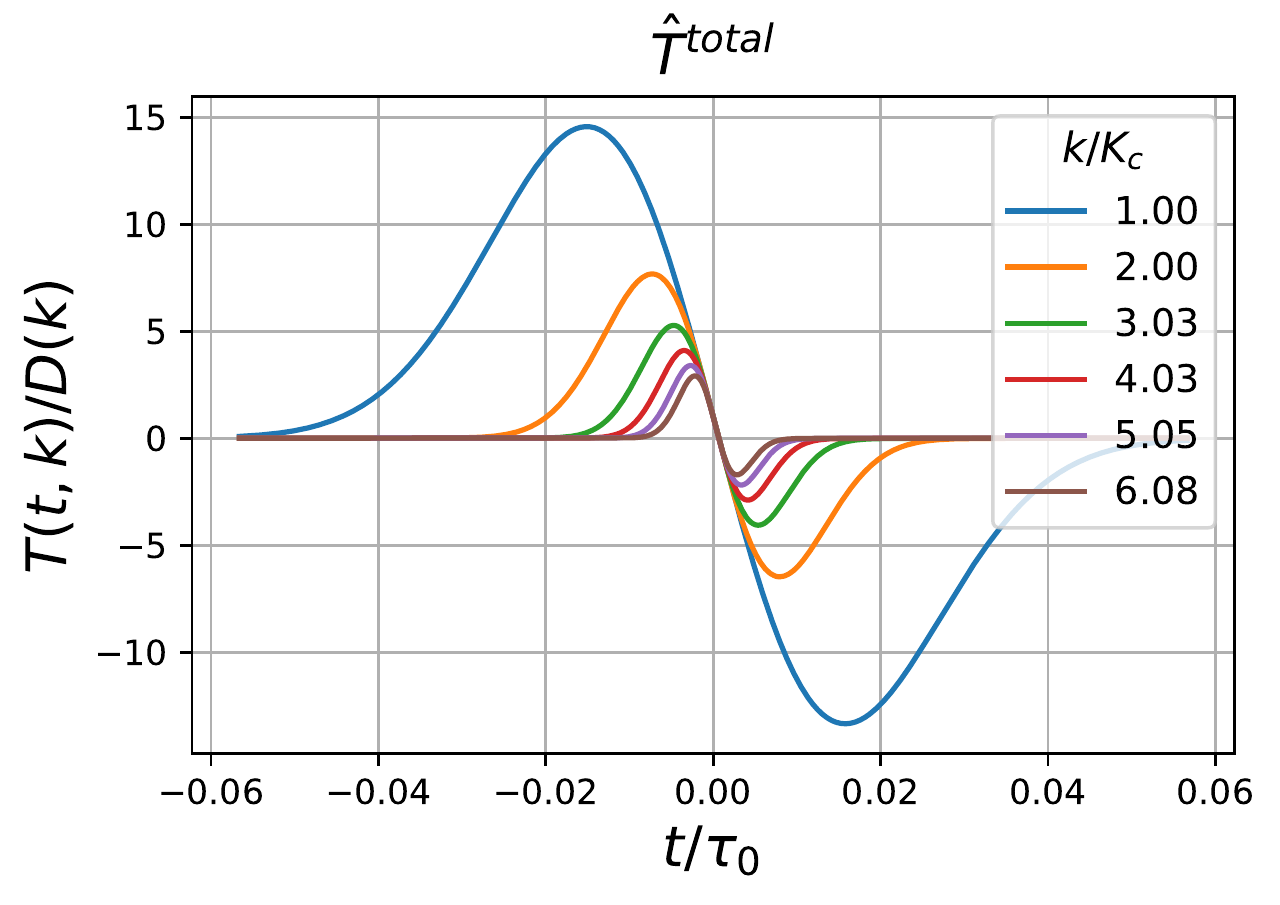}
	\includegraphics[width=0.75\linewidth]{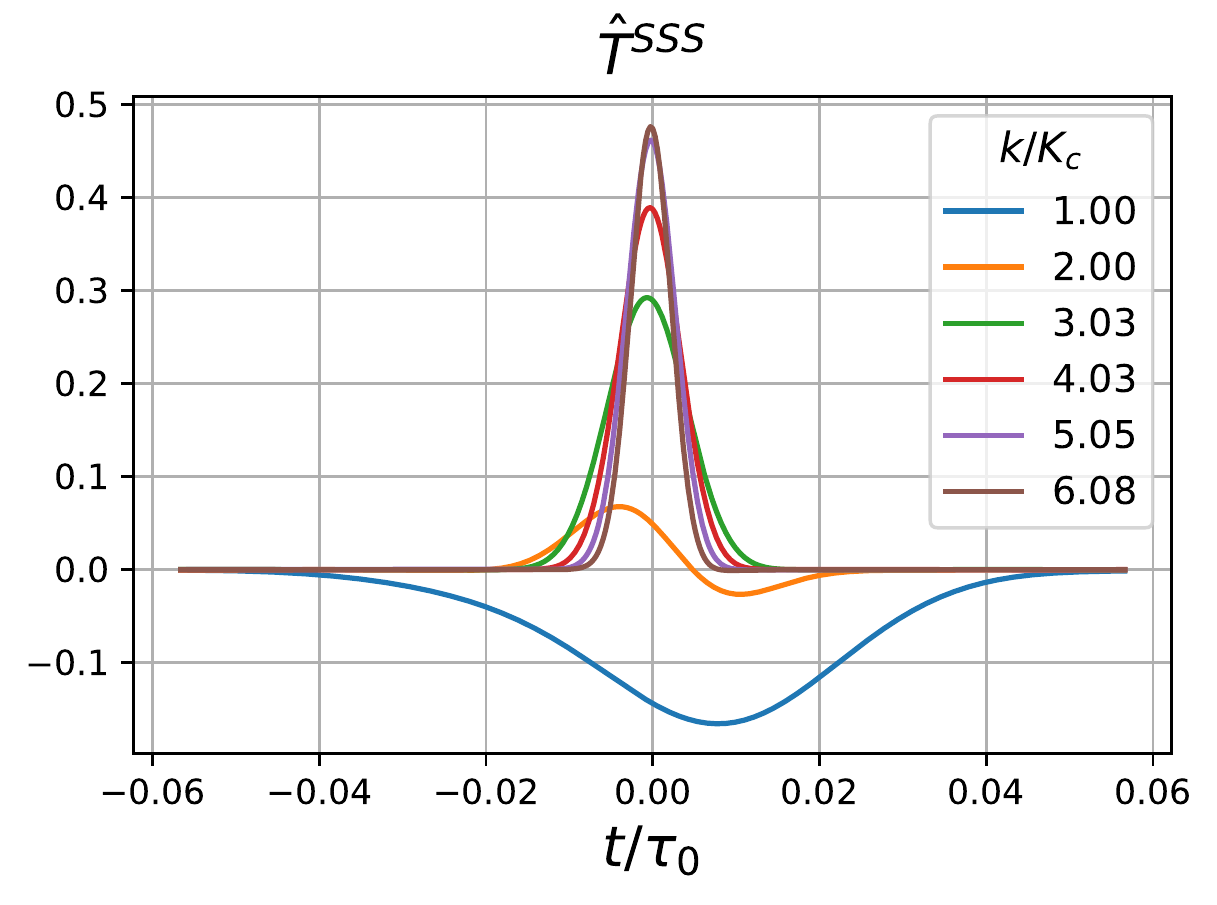}
	\caption{The advection-velocity spatio-temporal correlation function \(\hat{T}(t,\vec{k})\) versus time at selected values of wavenumbers \(k/K_c\):  total one (top panel),  small scale one $\hat{T}^{SSS}$ (bottom panel). The curves are normalized by the spectral dissipation rate \(\hat{D}(k) = \nu k^2 C^{(2)}(0,k)\).}
	\label{fig:tvst}
\end{figure}
The results for the time dependence of \(\hat{T}\) at different wavenumbers
\(k \gtrsim K_c\) are shown in  Fig.~\ref{fig:tvst}. One observes that the
total advection-velocity correlation function \(\hat{T}\) (top panel of 
Fig.~\ref{fig:tvst}) is not symmetric  with respect to the time origin and takes
negative values, in qualitative agreement with the form of the 
Eq.~\eqref{eq:tnonfiltered}. However, the term $\hat{T}^{SSS}$, which
only contains contributions from small scale modes to the correlation
function \(\hat{T}\), significantly changes shape (bottom panel of
Fig.~\ref{fig:tvst}). 

\par For the wavenumbers close to the cut-off wavenumber \(K_c\), the curves
are affected by the filter. To explain this, one should recall that at zero
time delay \(\hat{T}^{SSS}(t=0, \vec{k})\) is equal to the local nonlinear
energy transfer between small scales modes. At wavenumbers close to the
filter cut-off \(K_c\), some spectral modes participating in the local
energy transfers are suppressed by the filter. Thus, the modes close to the
filter cut-off transmit the energy to smaller scales, but they do not
receive energy from the nullified larger scales, which results in a negative
energy balance.  
\begin{figure}
	\centering
	\includegraphics[width=0.75\linewidth]{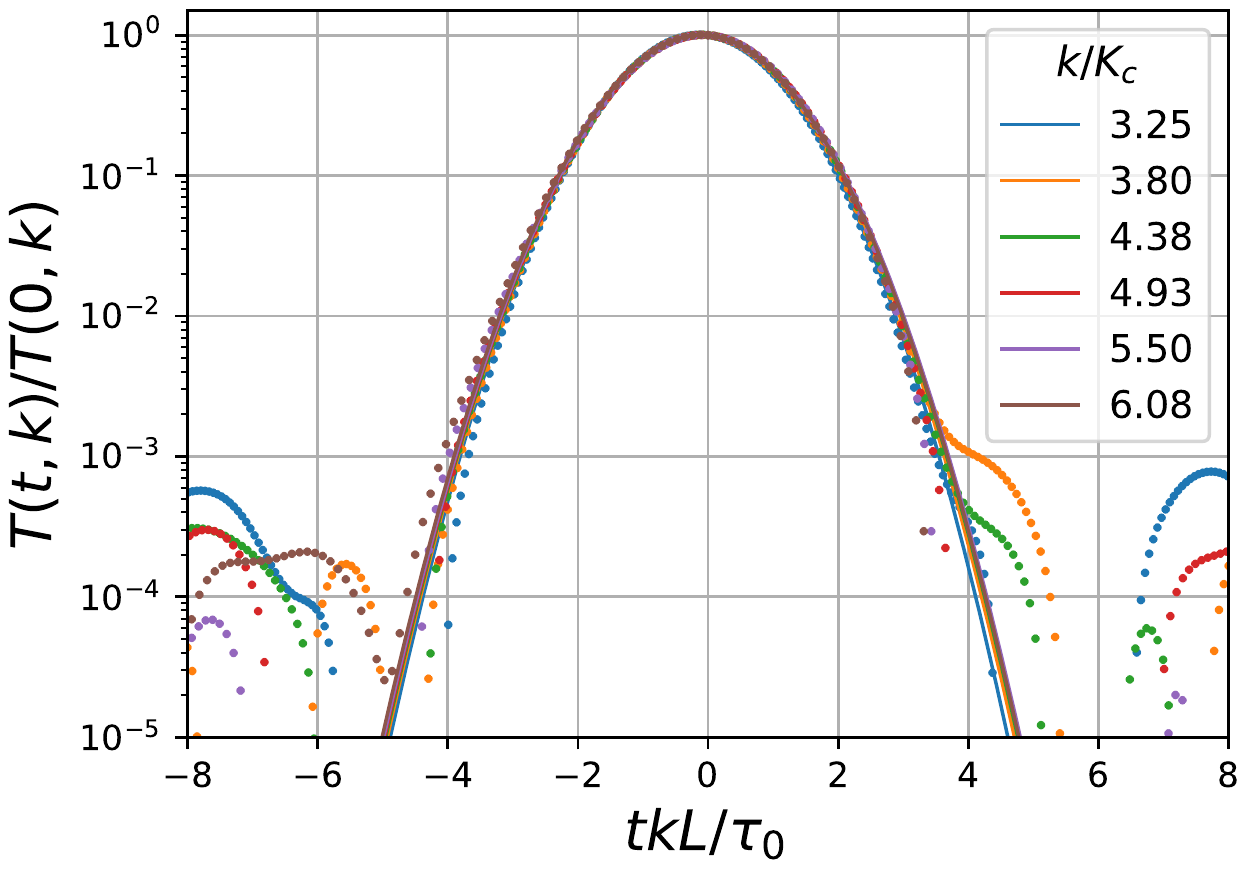}
	\caption{The small scale advection-velocity correlation function \(\hat{T}^{SSS}\) versus \(tk\) in semilog scale at various wavenumbers, for \(R_\lambda =160\), \(N=512\). The correlation functions are normalized by their value at \(t=0\).}
	\label{fig:tfiltvstcollapse}
\end{figure}
For the larger wavenumbers \(k\gtrsim 2K_c\), the curves deform towards the
expected Gaussian shape. This is further illustrated on 
Fig.~\ref{fig:tfiltvstcollapse}, where the correlation function 
\(\hat{T}^{SSS}\) is
plotted versus the scaling variable \(tk\) in semi logarithmic scale, inducing
a collapse of all the curves onto a single Gaussian. This is in plain agreement
with the theoretical result~\eqref{eq:theory3pcorr} for the three-point
correlation function. This behavior is very similar to the one for the
two-point correlation function presented in Fig.~\ref{fig:2pcorr}.
\begin{figure}
	\centering
	\includegraphics[width=0.75\linewidth]{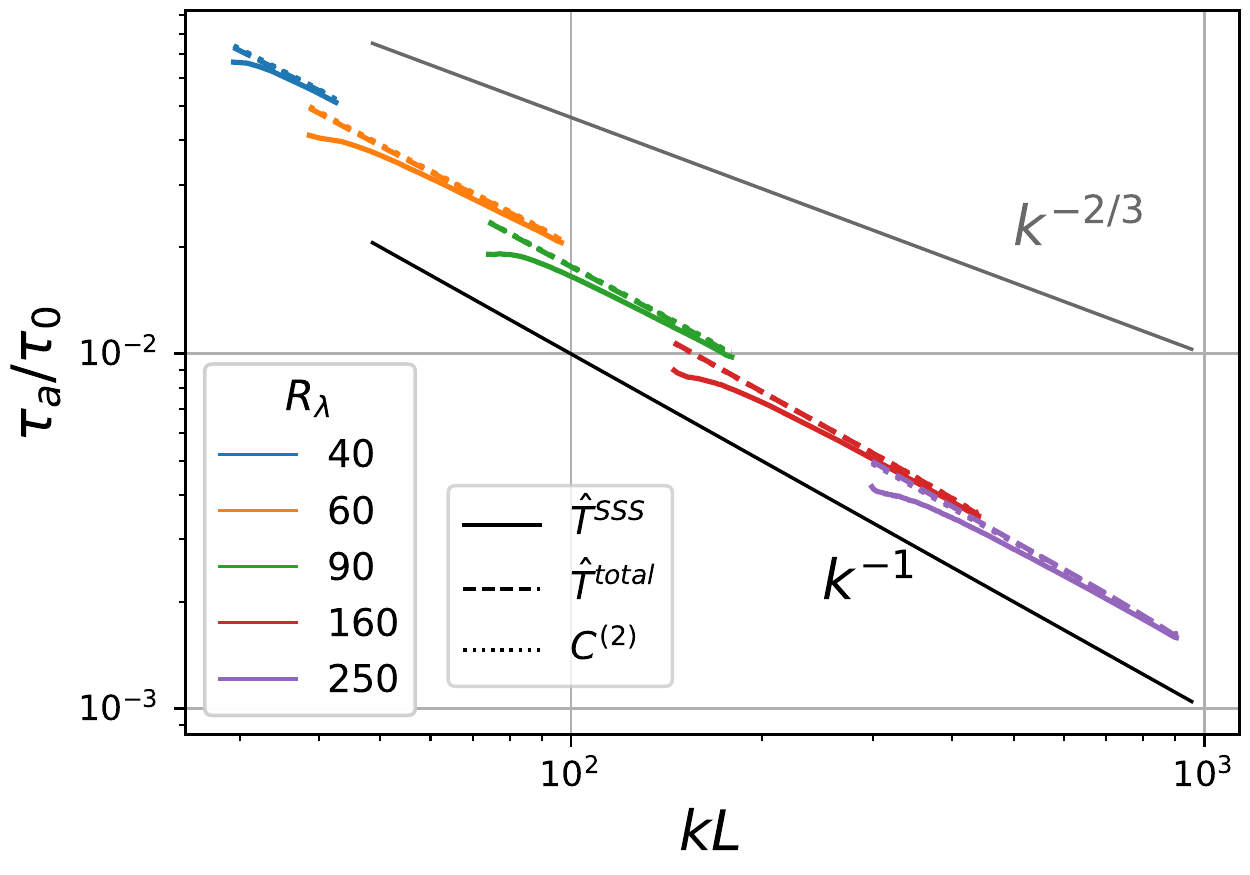}\\
	\includegraphics[width=0.75\linewidth]{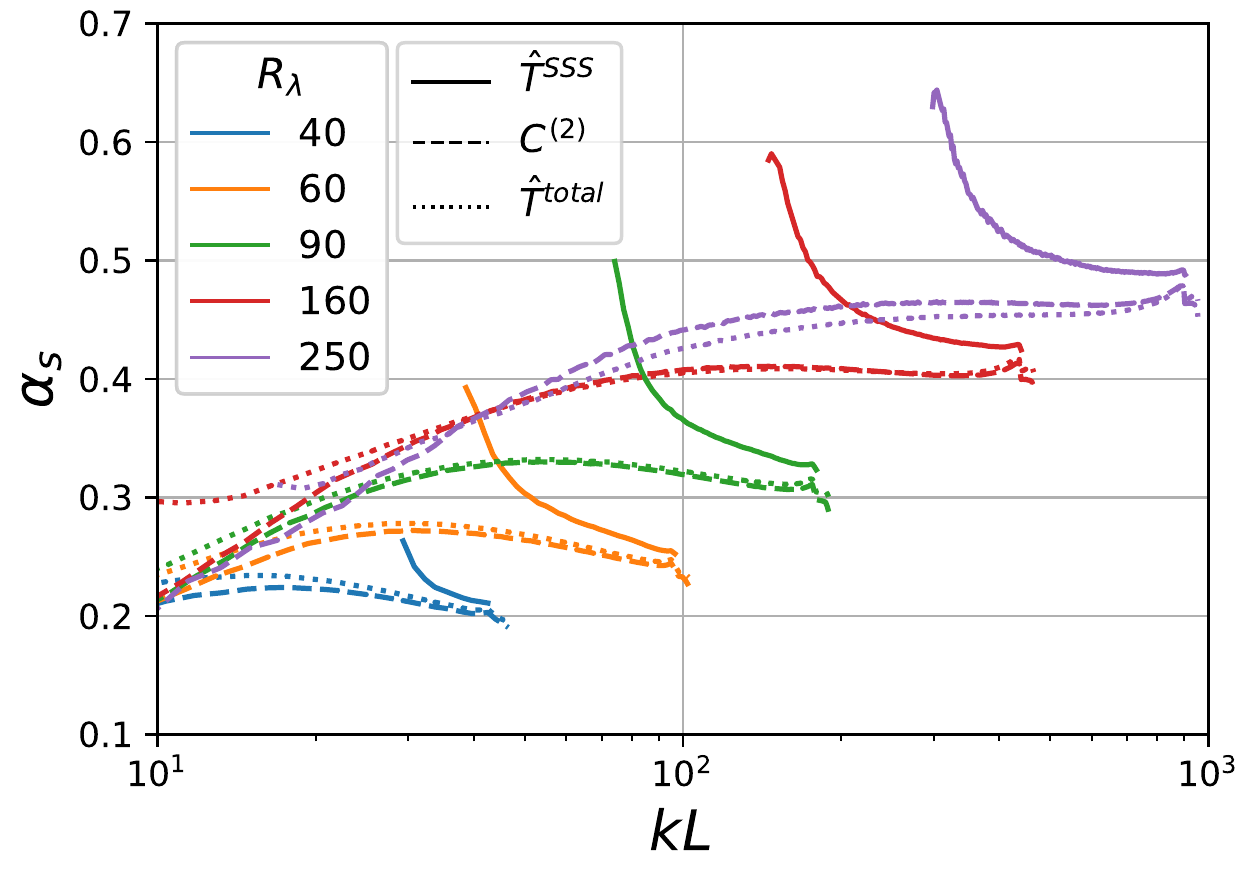}
	\caption{Numerical estimation of the parameter \(\tau_a\) (upper panel) and \(\alpha_S\) (bottom panel) 
	obtained from the small scale advection-velocity correlation \(\hat{T}^{SSS}\) (continuous lines), 
	compared with  the result for the two-point correlation function \(C^{(2)}\) from Fig.~\ref{fig:alphasvsk}
	 (dash-dotted lines). Both estimations converge to a similar value, as expected from the theory. 
	 The result of the fitting for the total advection-velocity correlation \(\hat{T}\) 
	 is also indicated with dotted lines for completeness.}
	\label{fig:tfiltfitavsk}
\end{figure}

\par 
We fit the curves obtained  
for the advection-correlation function \(\hat{T}\) with a function of
the form of Eq.~\eqref{eq:tnonfiltered}:
\begin{equation}
	f(t) = c\left(1 - \frac{t}{\tau_b}\right)e^{-(t/\tau_a)^2}
	\label{eq:fitting_T}.
\end{equation}
where $\tau_a, \tau_b$ and $c$ are the parameters.
\par We find that both correlation functions \(\hat{T}\) and $\hat{T}^{SSS}$ 
accurately fit \eqref{eq:fitting_T}.
Moreover, we verify that the fitting parameter \(\tau_a\) for both
functions is proportional to \(k^{-1}\), as displayed in 
Fig.~\ref{fig:tfiltfitavsk} (upper panel), and in agreement with  
Eq.~\eqref{eq:theory3pcorr}. We estimate from this parameter the value of
the coefficient $\alpha_S$ in Eq.~\eqref{eq:theory3pcorrtt} as \(\alpha_S = \tau_0/(\tau_ak^2L^2)\). 
The result is shown in
Fig.~\ref{fig:tfiltfitavsk}. At sufficiently large wavenumbers, the values of
$\alpha_S$ extracted from the small scale function \(\hat{T}^{SSS}\) and from
the total \(\hat{T}\) are comparable. They also match with the value
obtained from \(C^{(2)}\), as predicted by the theory. The small discrepancy visible between the values of
$\alpha_S$ from $C^{(2)}$ and from \(\hat{T}^{SSS}\) could be attributed to a
loss of accuracy due to the decomposition: The magnitude of the filtered signal
is much weaker, so it is more sensitive to the noise due to numerical errors.
\begin{figure}
	\centering
	\includegraphics[width=0.75\linewidth]{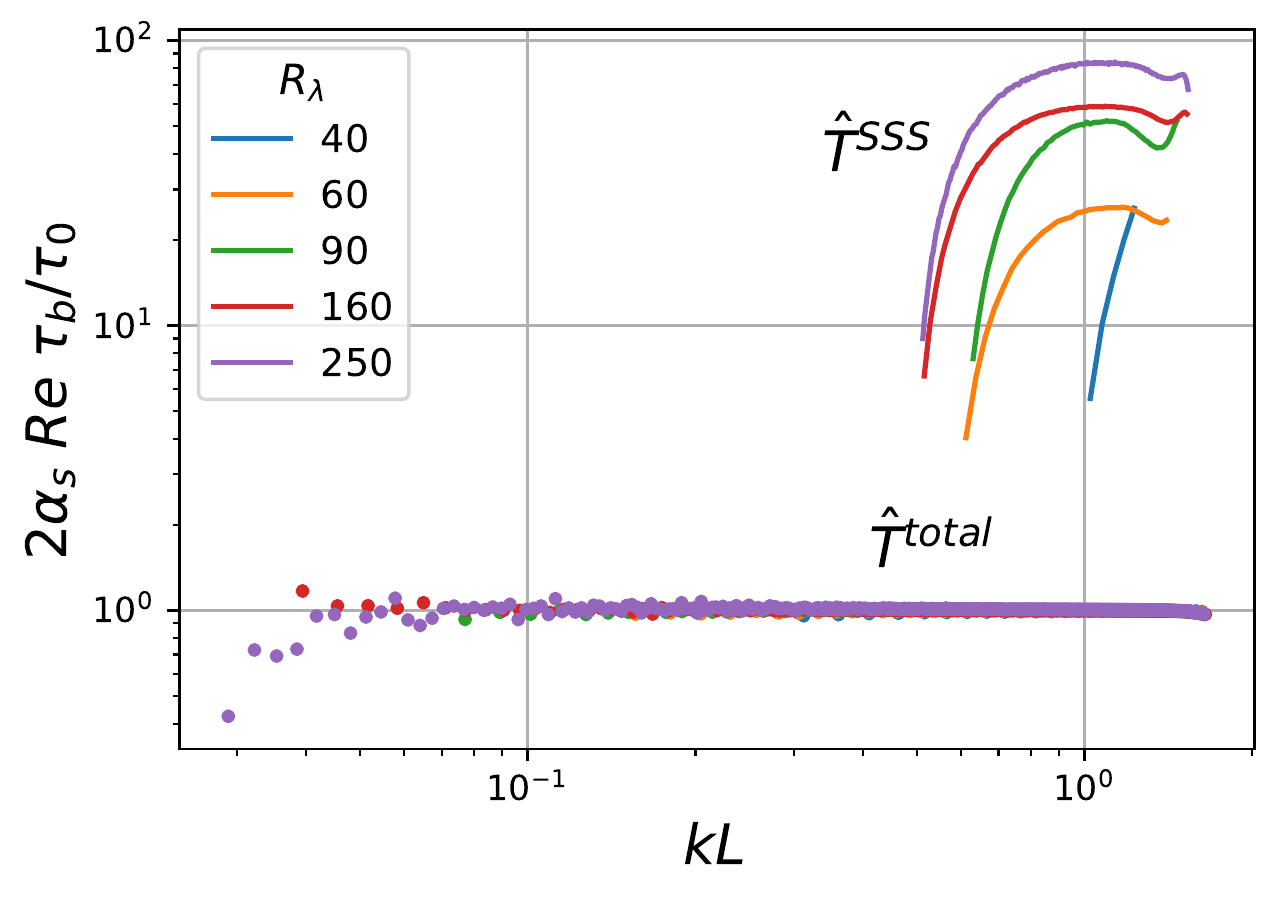}
	\caption{Dependence of the parameter \(\tau_b\) of the fitting function
Eq.~\eqref{eq:fitting_T} on the wavenumber for the small scale
advection-velocity correlation \(\hat{T}^{SSS}\) (continuous) and for the total
one \(\hat{T}\) (dotted lines). The values are normalized by $2\alpha_s
Re/\tau_0$ to enable comparison with the Eq.~\eqref{eq:tnonfiltered}.}
\label{fig:taub}
\end{figure}
\par Lastly, we examine the role of the parameter \(\tau_b\) in the fitting
function~\eqref{eq:fitting_T}. To do this, we can refer to the
Eq.~\eqref{eq:tnonfiltered} for the total correlation function $\hat{T}$,
which was obtained from the Navier-Stokes equation assuming that 
the two-point correlation function $C^{(2)}$ has a Gaussian shape.
Therefore, the fitting parameter time scale $\tau_b$ can be estimated as:
\begin{equation}
	\tau_b = \frac{\tau_0}{2\alpha_s Re}
	\label{eq:taub}
\end{equation} 
In the Fig.~\ref{fig:taub}, we show the dependence of the nondimensional parameter
\(2 \alpha_S Re \tau_b/\tau_0\) on the wavenumber. The values of $\alpha_s$ for
the normalization are taken from the fit of the two-point correlation function
$C^{(2)}$. As expected, for the total advection-correlation function
$\hat{T}$ the values from all simulations are in the vicinity of unity
independently from the wavenumber, which is consistent with the
Eq.~\eqref{eq:taub}. Besides, one can observe from the Fig.~\ref{fig:taub}
that for \(\hat{T}^{SSS}\) the non-dimensionalized parameter $\tau_b$ is at
least one order of magnitude larger than for
the total $\hat{T}$. This means
that for \(\hat{T}^{SSS}\) the time scale of the linear part $\tau_b$ of the
function~\eqref{eq:fitting_T} becomes much larger than the time scale $\tau_a$
of the Gaussian part. In other words, the Gaussian part decays fast and the
function already approaches zero before the slower linear part comes
into play, which results in the Gaussian-like shapes of $\hat{T}^{SSS}$ in 
the figures~\ref{fig:tvst} and \ref{fig:tfiltvstcollapse}. 
On the contrary, for the total
function $\hat{T}$ the time scale $\tau_b$ is smaller than $\tau_a$ and the
shape of the total $\hat{T}$ is dominated by the linear part at short times,
resulting in a non-symmetric shape. 
 
\par An interpretation of this result can be proposed based on the
identification of the advection-velocity correlation function $\hat{T}$ at
$t=0$ as the spectral energy transfer function. We have observed that a
significant part of the energy transfers between the small scales in 3D
turbulence occurs in spectral triads with participation of a large scale mode
as mediator (the term $\hat{T}^{SLS}$ in the 
decomposition~\eqref{eq:tdecompos}). The same conclusion can be found in 
Ref.~\onlinecite{vermaEnergyTransfersFluid2019,%
 domaradzkiLocalEnergyTransfer1990,%
 ohkitaniTriadInteractionsForced1992}. However, as discussed in Ref.
\onlinecite{aluieLocalnessEnergyCascade2009}, although 
these triads have significant individual contributions to energy transfer, they
are much less numerous than the fully local triads formed of small-scale modes
(the term $\hat{T}^{SSS}$ in the decomposition), because there
are fewer large-scale modes. In the
limit of large Reynolds numbers, the fully local triads become numerous and
dominate in the turbulent energy cascade. 

\par In addition, the detailed analysis
of the contributions in the decomposition \eqref{eq:tdecompos} shows that the
nonsymmetric behavior in time of the total correlation $\hat{T}$ is also
determined by the contribution of $\hat{T}^{SLS}$. The occurence of the
maximal and minimal values of the advection-velocity correlation $\hat{T}$
at non-zero time delays (see the top panel of the
Fig.~\ref{fig:tvst}) implies that there is some coherence between 
two small scale vortices simultaneously advected by a large scale, 
slowly varying, vortex. The origin of this coherence can be through an
alignment of turbulent stress and large scale strain rate.  The dynamics of the
alignment between time-delayed filtered strain rate and the
stress tensors, as well as its link with the energy flux between scales, 
has been recently analyzed in the
Ref. \onlinecite{ballouzTemporalDynamicsAlignment2020},
where the alignment also displays an asymmetrical behavior 
in time and is peaked at scale dependent time delays.
As the energy flux, which could be expressed as a
product of stress and strain rate, also represents a triple statistical moment
of the velocity field, it would be natural to expect that it exhibits a
temporal behavior similar to the advection-velocity correlation $\hat{T}$.

\par In the case of the purely small scale 
correlation function $\hat{T}^{SSS}$, the characteristic time scales of all modes in the triad are
comparable, and the mediator mode cannot impose any coherence on the
interacting modes, as all three modes decorrelate faster before any alignment
could occur. This results in the symmetric, close to Gaussian form of the small
scale correlation functions $\hat{T}^{SSS}$. Note that all the three modes in
$\hat{T}^{SSS}$ are still transported simultaneously by the random large scale
velocity field. This mechanism 
is the same random sweeping effect that is reponsible for the
Gaussian time dependence of $\hat{T}^{SSS}$ and of $C^{(2)}$. 

\par To conclude, the spatio-temporal correlation between the velocity and
advection fields constitutes a triple statistical moment easily accessible in
numerical simulations. The application of the scale decomposition to this
correlation is a necessary refinement to approach the regime of large
wavenumbers of the theoretical result and gives an insight into the statistics
of the three-point spatio-temporal correlation functions. We observe a Gaussian
with the same time and wavenumber dependence as in the theoretical prediction.
Moreover, this analysis provides also a nontrivial validation of the
theoretical result, which predicts the parameter $\alpha_S$ to be the same for
the two-point and the three-point correlations.

\subsection{Two-point spatio-temporal correlation of the modulus of the velocity}
\label{sec:2point-mod}

\par The numerical analysis of the two-point correlation function at large time delays represents a more challenging task,
as the values of the correlation functions become very low and are drowned 
into noise and numerical errors.
Moreover, it requires larger observation times,
and thus longer simulations and more computational resources. 
We did not succeed in resolving the large time regime
from our numerical data for the two-point correlation function, due to both the lack of statistics 
in the time averaging and the weakness of the signal, comparable with numerical errors.
 
\par However, in order to increase the amplitude of the signal,
we study the correlation function of the velocity modulus rather than  the real
part of the complex correlation function.  For this quantity, the large time
regime
indeed turns out to be observable, as we now report. We thus introduce the
connected two-point correlation function of velocity modulus in spectral space  
\begin{eqnarray}
\bar{C}_n^{(2)} (t,k) =  
\left< \lVert \hat{\vec{u}}({t_0}, \vec{k}) \rVert \  \lVert\hat{\vec{u}}({t_0} + t, \vec{k}) \rVert \right> - \nonumber \\
\left< \lVert \hat{\vec{u}}({t_0}, \vec{k}) \rVert \right> \left< \lVert\hat{\vec{u}}({t_0} + t, \vec{k}) \rVert \right>
\label{eq:def2pcorrLargeTimes}
\end{eqnarray}
with spatial and time averaging identical to 
Eq.~\eqref{eq:def2pcorrNumerical} 
$\left< ... \right> = \frac{1}{N_t} \frac{1}{M_n} \sum_{j=1}^{N_t}  \sum_{\vec{k} \in S_n} (...)$. 
This correlation function was computed in another set of simulations 
with larger width of the time window.
 
\begin{figure}
	\centering
	\includegraphics[width=0.75\linewidth]{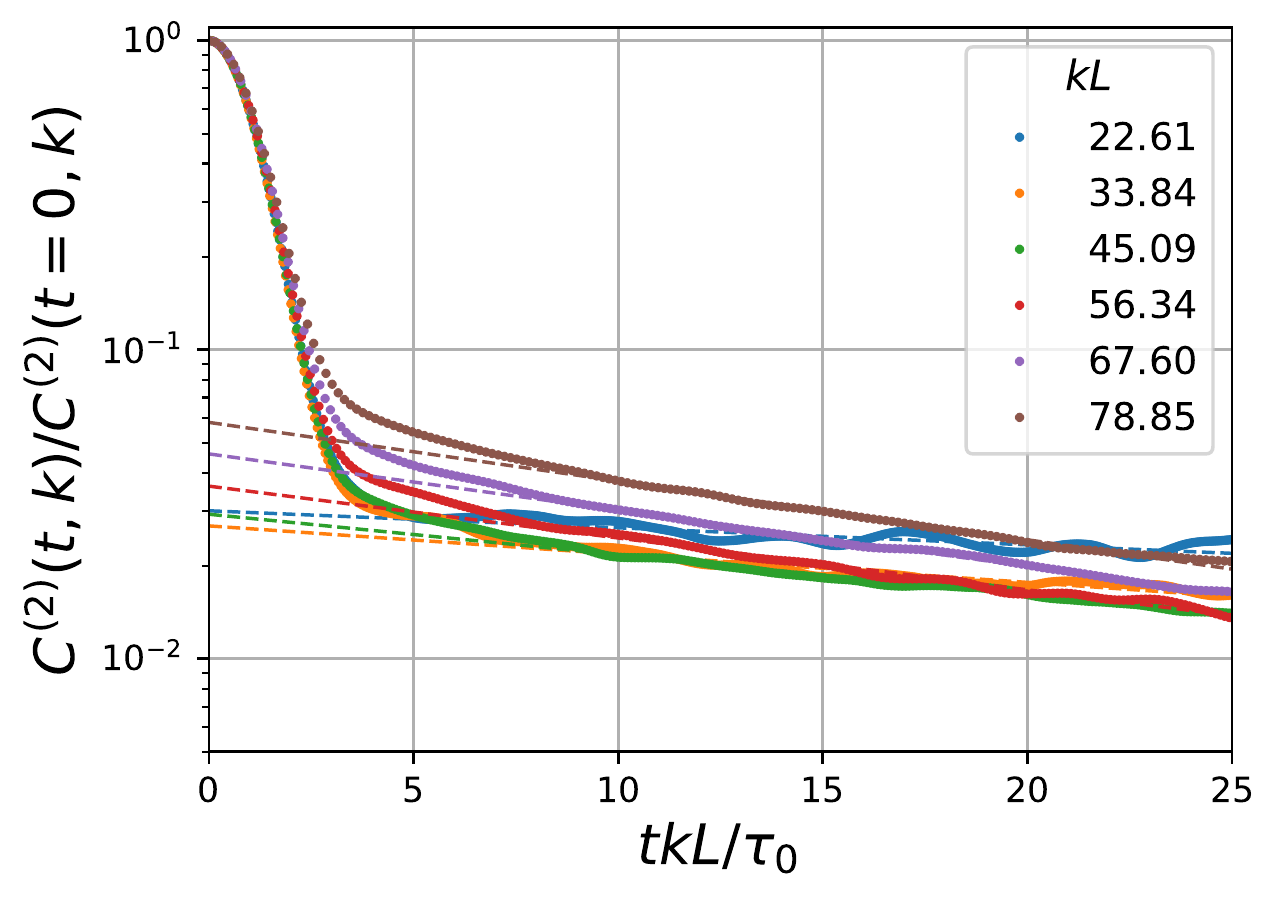}\\
	\caption{
Time dependence of the normalized two-point correlation function of the velocity norms \(\bar{C}_n^{(2)} (t,k)\) at $R_\lambda = 60$ for different wavenumbers \(k\) in semi-logarithmic scaling. 
The numerical data are denoted with dots, 
the exponential fit is denoted with the dashed lines.}
	\label{fig:2pcorrLarge}
\end{figure}
\par An example of the correlation function computed according to 
Eq.~\eqref{eq:def2pcorrLargeTimes} for $R_\lambda = 60$ is presented in 
Fig.~\ref{fig:2pcorrLarge}. Similarly to the two-point correlation studied in 
Sec.~\ref{sec:2point}, one observes at short time delays 
the Gaussian decay in time and the curves 
at different wavenumbers collapse in the $tk$-scaling. 
However, Fig.~\ref{fig:2pcorrLarge} reveals a crossover
to another regime at larger time delays: 
a slower decorrelation in time, that can visually be estimated as exponential. 
The curves at various wavenumbers do not collapse anymore in the horizontal scaling $tk$, 
and the slope of the correlation
function appears to be steeper for larger wavenumbers.  
\begin{figure}
	\centering
	\includegraphics[width=0.75\linewidth]{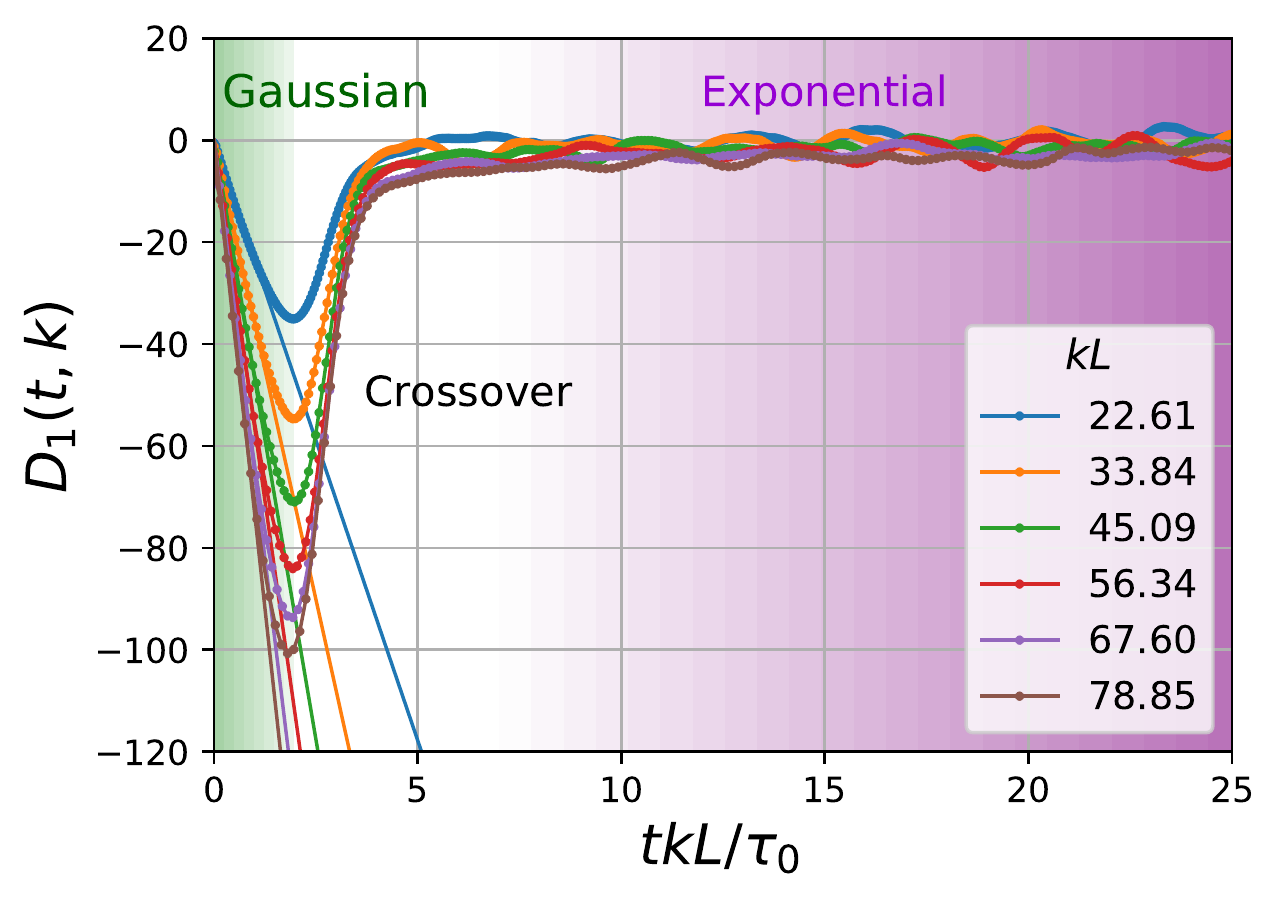}\\
	\caption{
	The normalized time derivative $D_1$ defined in Eq.~\eqref{eq:d1}
	 calculated numerically with the data from the simulation at $R_\lambda = 60$. 
	 The linear part of $D_1$, highlighted by green shades, corresponds to the Gaussian decay at small time delays,
	  and the approximately constant part of $D_1$, highlighted by purple shades,
	   corresponds to the exponential time correlation at large time delays.}
	\label{fig:crossover}
\end{figure}

\par
\begin{figure}
	\centering
	\includegraphics[width=0.75\linewidth]{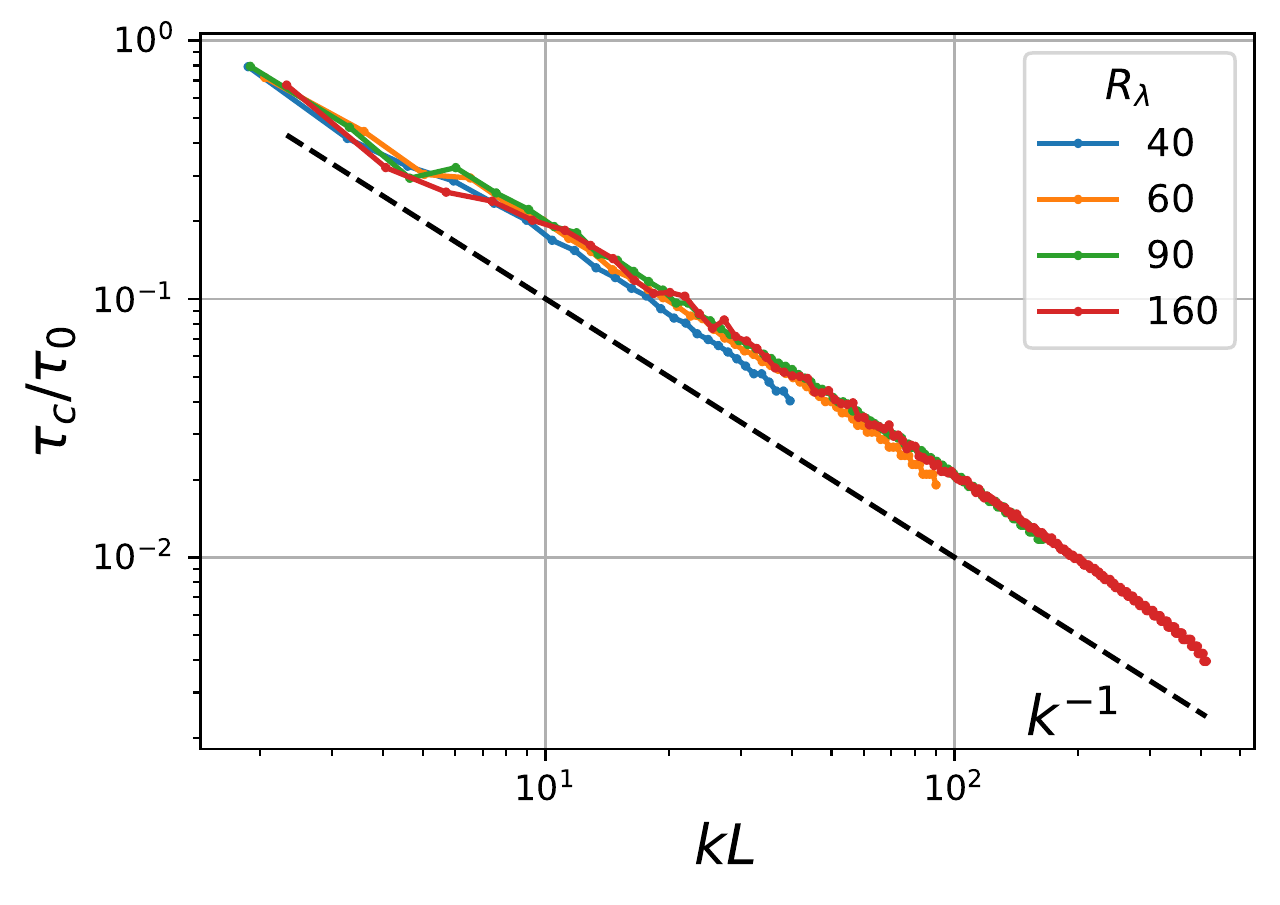}\\
	\caption{Crossover time for the two-point correlations of the velocity norms $\bar{C}^{(2)}_n$
	  between the small time and large time regimes as a function of  the wavenumber $k$, estimated
	   from the minimum of $D_1$. }
	\label{fig:taucross}
\end{figure}

 We compute the normalized time derivative of \(\bar{C}_n^{(2)} (t,k)\) in order to
study the transition between the two time regimes of the correlation function:
\begin{equation}
	D_1 (t, k) = \frac{\partial_t {\bar{C}_n^{(2)}} (t,k)}{\bar{C}_n^{(2)} (t,k)}.
	\label{eq:d1}
\end{equation} 
If the correlation function \(\bar{C}_n^{(2)}\) is a Gaussian, the time derivative
$D_1$ is simply a line with a slope equal to  $-2/{\tau_s^2}$, and if
the correlation function is an exponential function, the function $D_1$ becomes
a constant. The derivative $D_1$ is represented in Fig.~\ref{fig:crossover}
for $R_\lambda=60$. At small time delays, $D_1$ is a linear
function with a negative slope. 
It then displays a non-monotonous transition before
approximately reaching a constant value at large time delays. We can
define the crossover time delay  $t$ 
 as the location   of the minimum of the derivative $D_1$.
This crossover time at different Reynolds numbers is shown in the
Fig.~\ref{fig:taucross}. It depends on the wavenumber as $\tau_c\sim k^{-1}$. We checked that this $k^{-1}$ behavior does not depend on
the precise definition chosen for the crossover time.

\par Let us emphasize that the correlation function of the velocity norms 
introduced in 
Eq.~\eqref{eq:def2pcorrLargeTimes} is not related in any simple way to
the standard real part of the correlation function~\eqref{eq:def2point} 
computed theoretically
in the FRG approach. Moreover, as the phases play no role for these
correlations, the sweeping argument proposed in Sec.~\ref{sec:heuristic}
cannot explain this behavior. 
The decorrelation must ensue a priori from another
physical mechanism, yet to be identified.
However, the results of the numerical simulation 
show that the correlation of the velocity modulus
and the real part of the complex velocity 
correlation function at small time delays (the Gaussian decay) are similar, 
and exhibit close values for the characteristic decorrelation time. 
In addition, at large time delays the correlations of the velocity modulus
demonstrate a crossover to an exponential decay in time, analogous to the one expected for
the real part of the correlation function. 

While a complete understanding of these 
intriguing observations is lacking, some insight into the mechanisms at play in the 
regime of small time delays can be obtained from the expression valid to first-order 
in $t$
\[ \vec{u}({t_0} + t, \vec{r})=\vec{u}({t_0}, \vec{r}-\vec{u}({t_0}, \vec{r})t) 
-\vec{\nabla}p({t_0}, \vec{r})t+ {\cal O}(t^2), \]
where the second term which is required to enforce incompressibility involves the pressure 
satisfying the Poisson equation 
\( -\triangle p({t_0}, \vec{r}) = {\rm tr}[(\vec{\nabla}\vec{u}({t_0}, \vec{r}))^2]. \)
If one assumes that the $\vec{r}$-dependence can be ignored for the inner velocity 
field multiplied by $t,$ then this expression simplifies to 
$$\vec{u}({t_0} + t, \vec{r})=\vec{u}({t_0}, \vec{r}-\vec{u}({t_0}, \vec{0})t)+{\cal O}(t^2)$$
and one obtains $\hat{\vec{u}}({t_0} + t, \vec{k})
=e^{-it\vec{k}\cdot\vec{u}({t_0}, \vec{0})t}\hat{\vec{u}}({t_0}, \vec{k}),$ so that 
sweeping is represented by a pure change of phase of the Fourier mode. However, it is 
clearly inconsistent to neglect the $\vec{r}$-dependence of $\vec{u}({t_0}, \vec{r})$
in one instance and not in the other. Thus, the effects observed in Fig.~\ref{fig:crossover} must presumably be 
due to the spatial inhomogeneity of sweeping and the associated
long-range pressure forces arising from incompressibility, which decorrelate the moduli of the Fourier velocity amplitudes.
 If one furthermore plausibly assumes that the correlation $\bar{C}_n^{(2)}(t,k)$ 
is a maximum at $t=0$, then analyticity in $t$ requires in the regime of small time delays that   
$      \bar{C}_n^{(2)}(t,k) \doteq \bar{C}_n^{(2)}(0,k)(1- t^2/\tau^2_k ) $
for some parameter $\tau_k$ with units of time and then immediately 
$$                        D_1(t,k)  \doteq 1- \frac{2t}{\tau^2_k},  $$
as observed in Fig~\ref{fig:crossover}. These considerations do not explain 
the detailed observations, neither the $k$-dependence of $\tau_k$ nor the 
exponential decay in the regime of long time lags, but they do suggest some 
possible relevant physics for future theoretical and empirical exploration.  

\par Interestingly, a very similar behavior has been observed in the air jet experiments described 
in Ref. \onlinecite{poulainDynamicsSpatialFourier2006}. In these experiments, the temporal decay
of the two-point correlation function of the amplitude of the vorticity field
is measured, and it displays a crossover from a $tk$ Gaussian decay to a
slower exponential one. The crossover time
between these two regimes  is found to scale as $k^{-1}$ as observed in our
simulations \cite{Baudet2020}.

\section{Summary and Perspective}
\label{sec:conclusion}

In this paper, we use DNS to study
the spatio-temporal dependence of two-point and three-point correlations 
of the velocity field in stationary, homogeneous and isotropic turbulence. 
The motivation underlying this work
is to test a theoretical result obtained within the FRG framework, 
which gives the exact leading term at 
large wavenumbers of the spatio-temporal dependence of any $n$-point 
correlation function of the velocity field 
\cite{tarpinBreakingScaleInvariance2018}. 
This result establishes that the two-point correlation 
function decays as a Gaussian in the variable $tk$ 
(or $|\sum_i t_i \vec {k_i}|$ for a $n$-point correlation) 
at small time delays $t_i$, 
while at large time delays, the decorrelation slows down to a simple exponential in $t_i$.
While these results can in fact be interpreted quite simply by extending the
analysis of the random sweeping effect, following the original arguments by
Kraichnan, they are endowed through the FRG calculation with a rigorous and
very general expression.  In particular, these expressions show that for any
fixed time delays, the correlation function as
a function of any wavenumber is always Gaussian.  Furthermore, the
multiplicative constant in the exponential is the same for all the Gaussian
decays and all the exponential decays as well, independently from the
order~$n$.

In the small time regime that we could access via DNS of the two-point and
three-point correlation functions with an equal time delay, our
numerical data confirm the theoretical prediction with great
accuracy. 
In particular, we verify that the prefactors of time are  proportional
to $k^2$ (or $|\vec k_1+\vec k_2|^2$) and the numerical constants at small time
delays are indeed equal for the two-point and three-point correlations.
Furthermore, our analysis provides a deeper insight into  the range of validity
of the theory. All the theoretical results discussed here are derived under the
assumption that all the  wavenumbers (and their partial sums) are large. From
the DNS data, we estimate the range of $k$ where this condition is fulfilled
and show that it corresponds to the range where the direct energy transfer from
the forcing modes is negligible.  For the three-point correlations, we
show that once the small wavenumbers $k <K_c$ are removed through 
an appropriate decomposition, the theoretical prediction is precisely
recovered.
   
Our analysis of the correlation function of the modulus of the velocity
shows a very similar behavior as the one expected for the
velocity itself, although the theoretical results
do not apply in this case. It would be desirable to understand the main
physical mechanism at play for the decorrelation of the modulus, which cannot
be attributed to convective dephasing. 
This calls for further theoretical developments.
On the numerical side, it would be interesting to extend this analysis to
higher-order correlations, and for more general configurations in time 
(since our approach restricts to equal and short time delays 
for the three-point correlations). 
A particularly challenging task is the access to the long-time regime.
This would of course require important computing resources.
The understanding of the temporal correlations for passive scalars in turbulent
flows is also very important for many applications. This is work in progress.
   
\begin{acknowledgments}
We would like to thank C. Baudet for fruitful discussions and for
presenting us his experimental data. This work received
support from the French ANR through the project NeqFluids 
(grant ANR-18-CE92-0019). 
The simulations were performed using the high performance computing resources
from GENCI-IDRIS (grant 020611), {and the GRICAD infrastructure
(https://gricad.univ-grenoble-alpes.fr), which is partly supported by the
Equip@Meso project (reference ANR-10-EQPX-29-01) of the programme
Investissements d'Avenir supervised by the Agence Nationale pour la Recherche.}
GB and LC are grateful for the support of the 
Institut Universitaire de France.
\end{acknowledgments}


%

\end{document}